\documentclass[
twocolappendix,
twocolumn,
trackchanges,
amsmath
]{aastex702}

\usepackage{CJKutf8}
\usepackage{physics}
\usepackage{amssymb}
\usepackage{enumerate}
\usepackage{bm}
\usepackage{xspace}
\usepackage{xcolor}

\newcommand{\GMUNU}{\texttt{Gmunu}\xspace}
\newcommand{\WH}{\texttt{Weakhub}\xspace}

\newcommand{\corr}[1]{}
\newcommand{\strk}[1]{}

\graphicspath{{./}{figures/}}

\begin{document}

\title{Spinning down neutron-star merger remnants with the Tayler–Spruit dynamo:\\ Global simulations reveal the formation of massive disks and neutron-rich ejecta}

\author[0000-0003-3453-7394]{Harry Ho-Yin Ng}
\email[show]{hng@caltech.edu}
\affiliation{TAPIR, Mailcode 350-17, California Institute of Technology, Pasadena, CA 91125, USA}

\author[0000-0002-0491-1210]{Elias R. Most}
\email{emost@caltech.edu}
\affiliation{TAPIR, Mailcode 350-17, California Institute of Technology, Pasadena, CA 91125, USA}
\affiliation{Walter Burke Institute for Theoretical Physics, California Institute of Technology, Pasadena, CA 91125, USA}
\begin{abstract}
Magnetic-field amplification and angular momentum (AM) transport critically shape the secular evolution, lifetime, and electromagnetic signatures of binary neutron-star merger remnants. While the magnetorotational instability can operate in the outer negative-shear regions of the neutron-star remnant and accretion disk, the positive-shear, stably stratified core may instead be susceptible to the Tayler–Spruit dynamo. We present the first global, long-term general-relativistic neutrino-radiation magnetohydrodynamics simulations of a neutron-star merger remnant incorporating the unresolved Tayler–Spruit dynamo through a new mean-field dynamo subgrid prescription. Our axisymmetric simulations starting from a realistic merger remnant show that the Tayler–Spruit dynamo is primarily active in high-latitude regions of the remnant core. The resulting Maxwell stresses redistribute AM on a spin-down timescale of a few hundred milliseconds, substantially flattening the core rotation profile and transferring mass and AM from the outer remnant into the disk. This produces a more massive, extended, and strongly magnetized disk with a low electron fraction, leading to substantially more neutron-rich ejecta. Our results demonstrate that {currently unmodelled} Tayler–Spruit dynamo action can qualitatively alter the rotational evolution, collapse prospects, disk formation, and multi-messenger signatures of long-lived neutron-star merger remnants.
\end{abstract}


\section{Introduction}\label{sec:intro}
Binary neutron-star (BNS) mergers are key multi-messenger sources \citep{Lattimer74,Eichler89,Metzger:2010}, producing gravitational waves (GWs)~\citep{Abbott2017_etal}, kilonovae powered by the radioactive decay of ejecta undergoing r-process nucleosynthesis \citep{Kasliwal2017,Kilpatrick2017,Drout2017,Kasen2017,Villar2017},
 X-ray and radio emissions~\citep{Margutti2017,Mooley2018,Mooley2018b} from a structured jet and cocoon \citep{Gottlieb2018,Wu2019}, and short gamma-ray bursts (sGRBs)~\citep{FermiLat2017,Abbott2017d}.

The lifetime of the remnant determines the onset of sGRBs, the secular ejecta mass
loss, and the kilonova properties ~\citep{Hotokezaka2013,Dietrich:2015b,Radice2016,Bovard2017,Lehner2016,Radice2018a,Bernuzzi2020,Nedora2019,Nedora2020,Zappa2023}.
While various parameters and initial
conditions control the remnant lifetime~\citep{Gill2019,
Murguia-Berthier2020,Beniamini:2021tpy}, among the many factors, magnetic fields play a crucial role~\citep{Margalit2022}.
General-relativistic magnetohydrodynamic (GRMHD) simulations have shown that
the Kelvin--Helmholtz instability (KHI) \citep{Price06,Kiuchi2015a}, magnetic winding \citep{Shapiro00}, the
magnetorotational instability (MRI) \citep{Balbus1991}, and large-scale dynamo processes can
amplify the magnetic field, modify the field topology, and affect the remnant
dynamics {~\citep{Rasio99__,Price06,Giacomazzo:2014a,Kiuchi2015a,Ruiz2016,Ciolfi2019,Aguilera-Miret2020,Palenzuela_2022PRD,Chabanov2022,Celora2025,Fields2025,Gutierrez2025,
Kalinani2025,Neuweiler2025b,Rainho2025,Ng2026,Pais:2026ork,Gutierrez:2026ngt,Kiuchi2026}}.
A persistent difficulty in constraining the remnant lifetime is that the
relevant magnetic-amplification mechanisms and the corresponding angular
momentum (AM) transport are often unresolved {or not fully captured} in global simulations. {This is because effective (primarily neutrino-driven) viscous scales are often orders of magnitude smaller than the dynamical length-scales of the merger processs} \citep{Guilet2015,Skoutnev2024}. This has motivated large-eddy, subgrid, and mean-field
approaches for unresolved turbulent amplification, dynamo action, and AM
transport~\citep{Sadowski2015,Shibata2017b,Radice2017,Miravet2022,
Palenzuela_2022PRD,Duez2024a,Zhou2025,Cook:2026tum}, or
viscous-hydrodynamic models,
particularly for long-term simulations~\citep{Duez2004b,Fernandez2018,Fujibayashi2020b,Chawhan2025}.
For a long-lived postmerger remnant, the rotation profile remains differential
regardless of the initial conditions and equation of state
(EOS)~\citep{Hanauske2016,Uryu2017,Cassing2024}. One channel responsible for AM
transport in the remnant is the MRI-driven $\alpha\Omega$ dynamo \citep{Kiuchi2023}, but it
operates only in negative-shear regions, i.e., regions where the angular
velocity decreases outward, such as the outer remnant and disk \citep{Balbus1991}.

In regions where the shear profile is increasing, i.e., where the MRI cannot operate, another potential AM transport channel has been proposed: the
Tayler--Spruit (TS) dynamo. In this mechanism, non-axisymmetric modes of the
Tayler instability (TI)~\citep{Tayler1973}, a pinch-type instability of toroidal
magnetic fields, regenerate the poloidal magnetic field, while differential
winding amplifies the toroidal component, forming a self-sustained dynamo cycle~\citep{Spruit2002} in which the winding naturally transports AM \citep{Shapiro00}.
Inside stars, the TS dynamo is expected to operate in differentially rotating, stably stratified interiors and contribute to the spin-down of stellar cores~\citep{Spruit2002,Fuller:2019sxi,Ma:2019cpr}.
In proto-neutron stars spun up by fallback accretion, idealized models suggest that the TS dynamo could amplify magnetar-strength interior magnetic fields~\citep{Barrere2024}.
Idealized global MHD simulations of stably stratified spherical shells have begun to demonstrate self-sustained TS dynamo action and the associated AM transport~\citep{Petitdemange2023,Barrere2026b}.
Because resolving this process remains computationally infeasible, stellar-evolution models generally represent it as an effective viscosity in the AM equation~\citep{Paxton2015,Paxton2018,Fuller2019,Skoutnev2025}.

The TS dynamo has also been proposed to operate in the positive-shear core of BNS merger remnants, where it may efficiently amplify the magnetic field and transport AM, potentially shortening the remnant lifetime~\citep{Margalit2022,Reboul-Salze2025}. As a result, novel approaches to effectively modeling the TS dynamo in neutron-star merger calculations are needed.

We therefore present such an approach: the first global GRMHD simulations of a long-lived BNS merger remnant using a subgrid mean-field prescription to model the unresolved TS dynamo (Sec. \ref{sec:methods}) and quantify its impact on magnetic-field amplification, AM transport, secular evolution, and the emitted signals of long-lived remnants (Sec. \ref{sec:results}).
\section{Methods}\label{sec:methods}

\subsection{Tayler-Spruit dynamo}\label{sec:TSI}

In a background differentially rotating fluid
with angular velocity $\Omega(R)$, where $R$ is the cylindrical radius, winding amplifies the toroidal magnetic field \citep{Shapiro00} with the corresponding relativistic
Alfv\'en angular frequency $\omega_A = \sqrt{\sigma_{\rm tor}/(h+\sigma)}/R$,
where $\sigma=b^2/\rho$ is the magnetization, $b^2$ is the comoving magnetic-field strength, $\sigma_{\rm tor}\approx \sigma$ is the toroidal magnetization, $\rho$ is the rest-mass density, and $h$ is the specific enthalpy.
Under certain conditions, the toroidal field can become Tayler-unstable to small-scale radial perturbations \citep{Tayler1973}, which then generate a radial magnetic field \citep{Spruit2002}.
Although the TI can also operate in convectively unstable
regions, the overturning of radial fields and turbulent mixing can disrupt the ordered winding. It is therefore commonly assumed that the TI operates in stably stratified regions, i.e., where $N_{\rm BV}^{2}>0$, with $N_{\rm BV}$ denoting the Brunt--V\"ais\"al\"a frequency.
Under the appropriate hierarchy of frequencies,
$\omega_A \ll  \Omega \ll N_{\rm BV}$,
a sustained TS dynamo cycle can develop~\citep{Spruit2002}.
The thermal or compositional gradients required for a non-zero $N_{\rm BV}$ constrain the dominant cylindrical-radial wavenumber, $k_R$, of the TI \citep{Skoutnev2024}.
Here, we approximate it as~\citep{Spruit2002,Fuller2019}
\begin{equation}
k_R\gtrsim k_N\equiv N_{\rm BV}/(R\omega_A),
\label{eq:k_R_gt_k_N}
\end{equation}
consistent with the adiabatic
buoyancy used in our simulations. Thermal diffusion would instead select the
slightly smaller wavenumber $k_\kappa$~\citep{Skoutnev2024}.

Following the prescription of \citet{Fuller2019}, the TS dynamo is self-limiting: the TI transfers energy from the large-scale toroidal field into nonlinear magnetic and velocity perturbations, which cascade to smaller scales and are dissipated by microscopic diffusion. {Saturation is reached in either model when dissipation of either the background or field perturbations balance the onset of the TI, which happens at \citep{Skoutnev2024},} 
\begin{equation}
    \left|
    \frac{B^{R}}{R B^{\phi}}
    \right|_{\rm sat}
    {\sim \frac{1}{k_R R} \simeq}
    \frac{\omega_{A, \rm sat}}{N_{\rm BV}},
    \qquad
    \omega_{A, \rm sat}
    \simeq |\Omega| \left(\frac{R |\partial_R \Omega|}{ N_{\rm BV}}\right)^{1/3}\,,
    \label{eq:BrBphi_sat}
\end{equation}
where $B^i$ is an Eulerian magnetic-field component,
$\omega_{A,\rm sat}$ is
the saturated Alfv\'en angular frequency, and
$R$ is the cylindrical radius.
These relations were originally derived for approximately spherical stellar interiors~\citep{Fuller2019,Ma:2019cpr}, although recent work has also been devoted to clarifying the TI under neutron-star conditions \citep{Skoutnev2024}. We adapt them to cylindrical geometry to describe the approximately cylindrical {remnant}~\citep{Reboul-Salze2025}. Including general-relativistic corrections through the relativistic Ledoux discriminant $C_{\rm L}$~\citep{Mueller2013}, $N_{\rm BV}$ is {calculated} as
\begin{align}
    N_{\rm BV}^{2}
    \equiv&
    -c^{2}
    \frac{\alpha C_{\rm L}}
    {\rho h\psi^{4}}
    \partial_{n_P}\alpha\,,
    \nonumber\\
    =&\,
    N_{T}^{2}+N_{\mu}^{2}\,,
    \label{eq:N_bv}
\end{align}
where we have introduced the $3+1$ lapse function $\alpha$ and the conformal factor $\psi$ \citep{Gourgoulhon2012}.
Here, we decompose it into thermal ($N_T$) and composition ($N_\mu$) contributions.
We recast it along the local outward pressure-normal direction while retaining the cylindrical shear and winding terms. This treatment is better suited to the oblate, rapidly rotating {remnant} than approximating the local effective-gravity direction as purely radial; see App.~\ref{sec:BVfreq} for more details.

In neutrino-opaque merger remnants, neutrino transport may relax both thermal and lepton-fraction perturbations on sufficiently small scales. A consistent diffusion-reduced buoyancy frequency would therefore require a coupled treatment of energy and lepton-number diffusion and their competition with the TI growth rate. We instead adopt the adiabatic Ledoux buoyancy frequency of the background remnant as a conservative estimate of the stabilizing stratification.

In a postmerger remnant, regions with $d\Omega/dR>0$ are located primarily in
the high-density core, where neutrinos are trapped \citep{Zappa2023}.  When the
toroidal field is weak in these regions, short-wavelength TI modes can be
strongly damped by neutrino viscosity {\footnote{Because moment-based transport with an analytical closure does not capture neutrino shear
viscosity\citep{Gavassino2024}, we estimate it analytically. Specifically, we
adopt Eq.~(A8) of \citet{Margalit2022}, which accounts for neutrino--nucleon
scattering at arbitrary neutrino degeneracy, rather than the non-degenerate
approximation of \citet{Guilet2015}.}, $\nu_\nu$, } on scales larger than the
neutrino mean free path.  We therefore impose an approximate {critical
toroidal-field strength}, $B^\phi_{\rm crit}$~\citep{Reboul-Salze2025}, below
which our TI subgrid model is deactivated:
\begin{equation}\label{eqn:TS_neutrino_cond}
|B^\phi|
>
B^\phi_{\rm crit}
\equiv
\frac{\sqrt{\rho (h+\sigma)}}{\psi^2}
\left(
\frac{\nu_\nu N_{\rm BV}^2|\Omega|}{R^2}
\right)^{1/4},
\end{equation}
which includes relativistic corrections.

The TS dynamo introduces an electromotive force $\mathcal{E}_\phi$. In cylindrical geometry, it is expressed as \citep{Barrere2022}
\begin{equation}
\mathcal{E}_{\hat\phi}
\sim
\left\langle
\delta v_z \delta B_{\hat R}
-
\delta v_R \delta B_{\hat z}
\right\rangle,
\label{eq:TS_EMF_phi}
\end{equation}
which depends on the radial/meridional fluctuations of the magnetic field, $\delta B$, and velocity field, $\delta v$, as approximated by the mean-field model.
{Hatted quantities denote the orthonormal physical components.}
When expanded in terms of the background magnetic fields {$\bar{B}_{\hat{i}}$}, $\mathcal{E}_{\hat\phi}$ generally contains contributions from nondiagonal components of the 
mean-field $\alpha$-effect $\alpha_{ij}$ tensor as well:
\begin{equation}
\mathcal{E}_{\hat\phi}
=
\alpha_{\phi\phi}\bar B_{\hat\phi}
+
\alpha_{\phi R}\bar B_{\hat R}
+
\alpha_{\phi z}\bar B_{\hat z}
+\cdots \,.
\label{eq:TS_EMF_general}
\end{equation}
We estimate the mean-field $\alpha$-effect of the TS dynamo by
projecting the saturated Tayler-mode perturbations onto the azimuthal component of the turbulent electromotive force $\mathcal{E}_{\hat\phi}$, retaining only the dominant $\alpha_{\phi\phi}$ contribution,
\begin{equation}
\mathcal{E}_{\hat\phi}
\approx
\alpha^{\rm TS}_{\phi\phi}\bar B_{\hat\phi},
\label{eq:alpha_TS_definition}
\end{equation}
where a hatted subscript denotes the orthonormal physical
component of the quantity, and $\bar{B}_{i}$ is the mean magnetic
field. As we will see, this simplification makes it straightforward to adopt {relativistic} mean-field dynamo approaches \citep{Bucciantini2012a,Shibata2021c,Most2023b}.
The
omitted terms act on the mean poloidal field, which is limited by
stratification to $\bar B_{\rm pol}/\bar B_{\rm tor}\lesssim
\omega_{A,{\rm sat}}/N_{\rm BV}\ll1$, as monitored by our local saturation condition, i.e., Eq.~\eqref{eq:kappa_tar}. Although the omitted coefficients may be large because of the mode anisotropy, these terms are suppressed during the initial phase and reach $\mathcal{O}(1)$ only near local saturation, where the subgrid model is quenched {(see Eq.~\ref{eq:kappa_tar} later)}.
The residual order-unity
uncertainty of the unresolved tensor structure is absorbed into
$\chi_{\rm TS}$; adoption of the full $\alpha_{ij}$ is left to future work.

In a stably stratified fluid, the radial {perturbation} of the Tayler mode is buoyancy-suppressed relative to the horizontal displacement. Using the Tayler-mode aspect ratios at local saturation from \citet{Fuller2019}, but adapting them to cylindrical rather than spherical-polar geometry, we take
\begin{equation}
\frac{\delta v_R}{\delta v_\perp}
\sim
\frac{\delta B_{\hat R}}{\delta B_\perp}
\sim
\frac{\omega_{A,{\rm sat}}}{N_{\rm BV}} .
\label{eq:TS_radial_leg}
\end{equation}
Thus,
\begin{equation}
\mathcal{E}_{\hat\phi}
\sim
\frac{\omega_{A,{\rm sat}}}{N_{\rm BV}}\,
\delta v_\perp \delta B_\perp,
\label{eq:TS_EMF_radial_leg}
\end{equation}
where $\delta v_\perp$ and $\delta B_{\perp}$ are
perturbations of velocity and magnetic field
perpendicular to the radial direction, respectively.
Here, the factor $\omega_{A,{\rm sat}}/N_{\rm BV}$ is the anisotropic projection factor associated with the buoyancy-suppressed radial leg.
{In the model by \citet{Fuller2019}, saturation involves the dissipation of the perturbation, and directly constrains (\citet{Fuller2019}, their Eqs.~(12), (15), and (16)),}
\begin{equation}
\frac{\delta B_\perp}{\bar B_{\hat\phi}}
\sim
\frac{\omega_{A,{\rm sat}}}{|\Omega|},
\qquad
\delta v_\perp
\sim
\frac{\omega_{A,{\rm sat}}}{|\Omega|}\delta v_A,
\qquad
\delta v_A
\sim
R\frac{\omega_{A,{\rm sat}}^2}{|\Omega|}.
\label{eq:TS_fuller_fluctuations}
\end{equation}
Therefore, by combining Eqs.~\eqref{eq:alpha_TS_definition}, \eqref{eq:TS_EMF_radial_leg}, and \eqref{eq:TS_fuller_fluctuations}, we set
\begin{equation}
 \kappa_{\rm TS}
 :=
{|\alpha^{\rm TS}_{\phi\phi}|}
=
\chi_{\rm TS}
\frac{R\,\omega_{A,{\rm sat}}^5}
{N_{\rm BV}|\Omega|^3},
\label{eq:kappa_TS_final}
\end{equation}
where we have added $\chi_{\rm TS} = 0.1$ as a parameter
for the dynamo efficiency that conservatively absorbs intrinsic model uncertainties {(see App.~\ref{sec:diff_chi} for for an assessment of the impact of $\chi_{\rm TS}$)}.

\subsection{Subgrid dynamo model}

Capturing a {sustained} TS dynamo cycle requires sufficient separation among the physical dissipation scales, the numerical viscosity scale, and the characteristic length scales of the growing TI modes, because the peak wavenumbers of the Tayler instability cannot be resolved in the magnetic Prandtl, $\rm{Pm} \sim 1$, regime (e.g., with uncontrolled numerical diffusion~\citep{Skoutnev2024}). In this work, to perform integrations for up to $0.5~\rm{s}$, we consider axisymmetric evolutions {(see also \cite{Fujibayashi2017b,Shibata2021c,Ng2024b})}. Since {axisymmetric} systems cannot sustain dynamo action~\citep{Cowling33}, we employ a combined subgrid model for both the TS and MRI dynamos, the latter of which is expected in the outer layers of the neutron-star merger remnant and inside the accretion disk \citep{Kiuchi2023}. As argued above for the TS dynamo and for the MRI by \citet{Most2023b}, we assume that, to leading order, the electric field in the fluid-comoving frame depends linearly on the comoving magnetic field,
$e^\mu = \kappa b^\mu$,
where $\kappa\propto |\alpha_{\phi\phi}|$ is a pseudoscalar proportional to the azimuthal component of the mean-field dynamo $\alpha$-tensor, $\alpha_{\phi\phi}$. We approximate this $\alpha$-effect as isotropic, with $\kappa\ll1$, and express an effective Eulerian electric field as \citep{Most2023b}:
$$
E^i_{\rm MFD} =
-\varepsilon^{ijk} v_j B_k +
\kappa
\left[
\left(1-v^2\right) B^i
+
\left(v_l B^l\right) v^i
\right]
+
\mathcal{O}\left(\kappa^2\right),
$$
where
$v^i$ is
Eulerian three-velocity,
$\varepsilon^{ijk}
= \tilde{\epsilon}^{ijk}{/(R\psi^{6})}$ 
and $\tilde{\epsilon}^{ijk}$ is
the Levi-Civita symbol.
The saturation of the dynamo cannot be self-consistently captured in {subgrid-dynamo simulations} and requires an effective criterion, which we supply as follows.
For the TS dynamo using Eq. \eqref{eq:BrBphi_sat}, we obtain the target saturation closure for the
relaxation equation for $\kappa$:
\begin{equation}
\begin{aligned}
\kappa_{\rm tar}
    &= \kappa_{\rm TS}\,\Delta_{\rm TS}, \\
\Delta_{\rm TS}
    &= \min\!\left[
        1,\,
        \max\!\left(
            0,\,
            1-
            \frac{\left|B^{R}/B^{\phi}\right|}
                 {\xi_{\rm TS}
                  \left|B^{R}/B^{\phi}\right|_{\rm sat}}
        \right)
    \right],
\end{aligned}
\label{eq:kappa_tar}
\end{equation}
where $\xi_{\mathrm{TS}}$ is a free parameter that controls the saturation level, and $\xi_{\rm TS} = 1$ is chosen for the fiducial model.

To compute $\kappa_{\rm TS}$, our model uses the saturated value $\omega_{A,\rm sat}$ rather than the instantaneous value $\omega_A$, thereby approximating the upper-bound effect of the TS dynamo. Varying the saturation parameter $\xi_{\rm TS}$ and efficiency parameter $\chi_{\rm TS}$ then allows us to span progressively weaker dynamo effects below this upper bound.

Apart from the stratification, {shear gradient,} and scaling conditions adopted from \citet{Fuller2019}, we apply physical limitations to the subgrid model, {incl. Eq. \eqref{eqn:TS_neutrino_cond}.}

We also consider an MRI-driven $\alpha\Omega$ dynamo, which is restricted to
regions with $d \Omega/ dR < 0$, i.e., the outer layers of the remnant and the accretion disk.
We employ the saturation factor of Eq.~(35) in
\citet{Most2023b} and
the calibration from \citet{Radice2020}, corresponding to
$\Delta_{\rm MRI} = \min[1,\max(0,1 - \sigma/\sigma_{\rm turb})]$ with $\xi_{\rm MRI} = 4$, where
$\sigma_{\rm turb}$ is saturation-level magnetization.
However, we replace the constant parameter $\kappa_{\rm MRI}$ used in \citet{Most2023b} with a more faithful prescription for the accretion-disk dynamics we intend to capture \citep{Sadowski2015}:
\begin{equation}
\kappa_{\rm MRI}
= \chi_{\rm MRI}
\left(\frac{2|R \partial_R \Omega|
H_{\rm eff}}{3}\right),
\label{eq:kappa_MRI}
\end{equation}
where the Keplerian angular velocity is replaced by
the local shear frequency $2R \partial_R \Omega/3$,
$\chi_{\rm MRI}=0.05$ is chosen~\citep{Sadowski2015}, and
$H_{\rm eff}$ is the scale height of the thermally pressure-supported matter.

\subsection{Numerical setup}\label{sec:setup}
We numerically solve the 2.5D axisymmetric general-relativistic {neutrino-}dynamo-magnetohydrodynamics equations in a conformally flat spacetime. The initial remnant is modeled using the fully nonlinear outcome of a numerical relativity calculation.
For this, we {first} use the outcome of a three-neutrino-species simulation of an irrotational binary neutron-star merger from \citet{Ng2024c} (see the setup therein), adopting the SFHo EOS~\citep{Steiner2013} and a total binary gravitational mass of $2.5~M_{\odot}$.
Spin-down of {this BNS merger remnant} will lead to a stable, uniformly rotating neutron star. 
The simulation was performed with the full-GR ideal-GRMHD neutrino-radiation code \texttt{FIL-M1} in three-dimensional (3D) Cartesian geometry~\citep{Most2019b,Musolino2023}. \texttt{FIL} is derived from the \texttt{IllinoisGRMHD} code \citep{Etienne2015}. In \texttt{FIL-M1}, we additionally impose a purely poloidal magnetic seed $2~\rm{ms}$ before merger, generated by the toroidal vector potential
$A_{\phi} = A_0 \max(0, p-0.01 p_{\rm max}),$
where $p$ is the pressure, $p_{\rm max}$ is the maximum pressure inside the binary stars, and $A_0$ is chosen such that the maximum poloidal-field strength is $B^{\rm pol}_{\rm max}\sim10^{14}~\rm{G}$ at the time of insertion.
We evolve the merger until $15~\rm{ms}$ postmerger ($\bar t_0 = 15~\rm{ms}$, where $\bar t = t - t_{\rm mer}$ is the postmerger time measured from the moment of merger $t_{\rm mer}$) and then perform a coordinate transformation and $\phi$-averaging of the 3D data.
{At this point, the mass-averaged magnetic-field strength is around $10^{15}~\rm{G}$.}
The hydrodynamic, magnetic-field, radiation, and metric quantities are transferred to another ideal-GRMHD neutrino-radiation code, \GMUNU, which employs the conformal-flatness condition in cylindrical geometry $(R,z,\phi)$~\citep{Cheong2021}, to continue the long-term axisymmetric simulation~\citep{Ng2024b}. The spurious oscillatory feature in the temperature profile caused by the $\phi$-averaging and gauge difference, as reported in \citet{Ng2024b}, is removed by remapping the temperature profile after metric initialization. We leverage the high efficiency of the axisymmetric approximation~\citep{Lam2025}, motivated by the nearly axisymmetric postmerger remnant after the damping of GW emission~\citep{Hanauske2016}, together with the {conformally flat} gravity scheme, which further improves efficiency while remaining highly accurate with respect to full general relativity~\citep{Ng2024b}.
{The metric variables are updated at intervals of} $\Delta t_{\rm met} \simeq 5\times10^{-1}~M_{\odot}$~\citep{Jiang2025}.

\begin{figure*}
    \centering
    \includegraphics[width=0.95\textwidth]{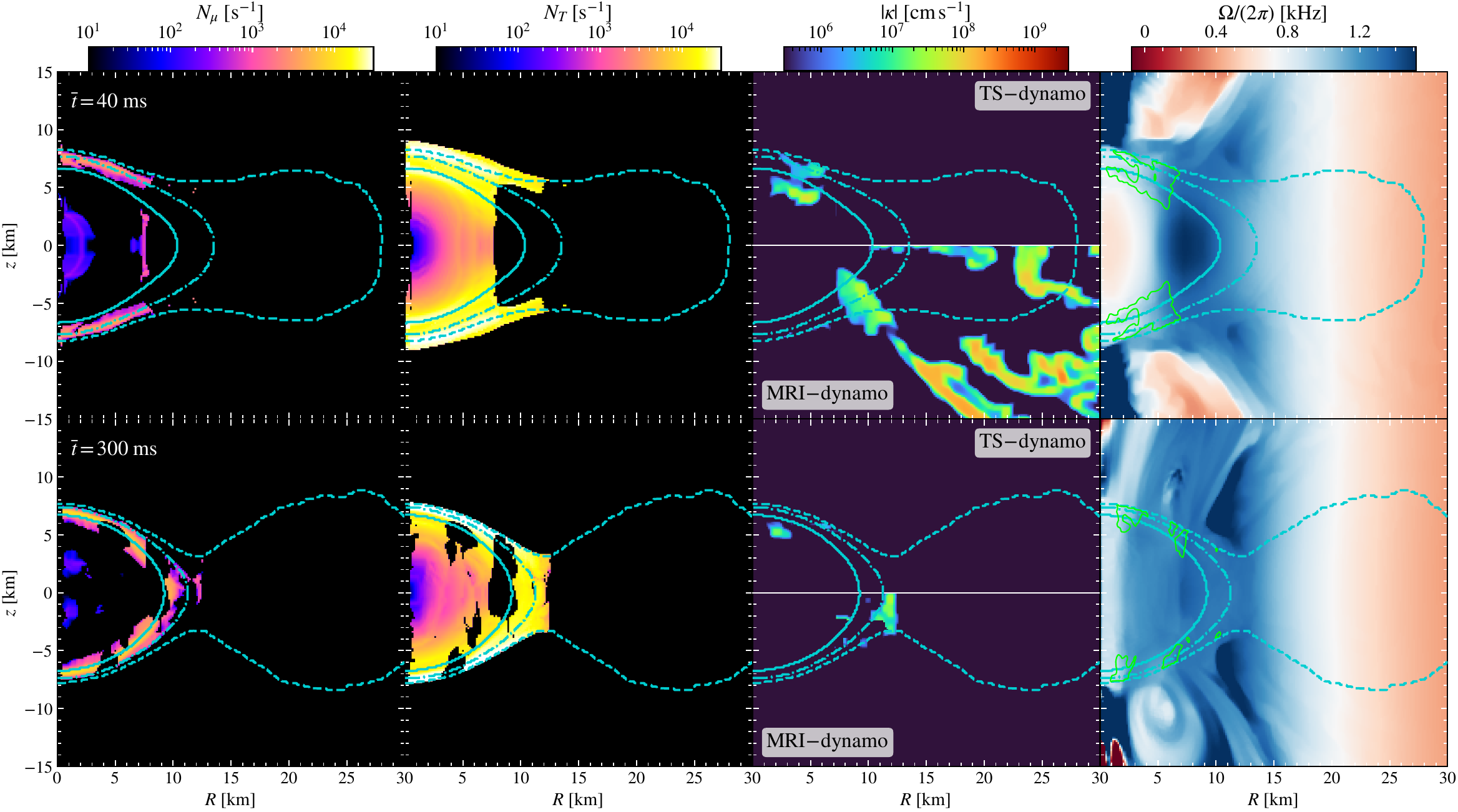}
    \caption{
    \textit{From left to right:} Meridional slices {of the fiducial ($\xi_{\rm TS}=1$) model} of the composition ($N_\mu$)
    and thermal ($N_T$) contributions to the Brunt--V\"ais\"al\"a frequency,
    the absolute value of the isotropic $\alpha$-effect coefficient, $|\kappa|$, and the rotational
    frequency, $\Omega/(2\pi)$. The snapshots are shown at
    $\bar t=40~{\rm ms}$ (top row) during the TS-dynamo growth stage and $\bar t=300~{\rm ms}$ (bottom row).
    In the $|\kappa|$ panels, the upper half-plane shows the contribution from the
    TS dynamo, while the lower half-plane shows that from the MRI dynamo.
    Cyan dashed, dash-dotted, and solid curves denote rest-mass density contours of
    $10^{12}$, $10^{13}$, and $10^{14}~{\rm g~cm^{-3}}$, respectively.
    Green contours mark regions where $|\Omega|=\Omega_{\rm TS}\equiv
    \sqrt{\omega_{A,\rm sat}N_{\rm BV}}$,
    where the saturated-state TI growth timescale equals the
    AM-transport timescale estimated from Eq.~(35) of
    \citet{Fuller2019}.
    {The regions enclosed by the green contours correspond to 
    $|\Omega|<\Omega_{\rm TS}$.} 
    }
    \label{fig:fig1}
\end{figure*}
For the \GMUNU continuation, we extend the computational domain to
$(R,z)\in[0,2500]\times[-2500,2500]~{\rm km}^2$.
The simulation employs eight refinement levels, with a finest
resolution of $150~{\rm m}$; the finest refinement box covers
$R\leq 33~{\rm km}$ and $|z|\leq 18~{\rm km}$.
The $\phi$-averaging can partially disrupt the magnetic-field topology and underestimate the toroidal magnetic energy because of cancellations. We therefore $\phi$-average the toroidal field using its root-mean-squared value, with the sign set by the winding-induced equatorial parity.

\GMUNU controls the magnetic-field divergence using a combination of
hyperbolic divergence cleaning and multigrid elliptic cleaning~\citep{Dedner:2002}.
The subgrid dynamo model is implemented using the scheme of \citep{Most2023b}.
Simulations adopt the Harten, Lax, and van Leer
approximate Riemann solver~\citep{Harten83},
the third-order piecewise-parabolic reconstruction method
~\citep{colella_1984_ppm}, and the IMEXCB3a time integrator
~\cite{Cavaglieri2015}.
For radiation, we employ the energy-averaged (grey) moment scheme~\citep{Cheong2023,Cheong2024b},
coupled to the neutrino microphysics library \WH~\citep{Ng2024a}
(see App.~\ref{app:micro} for details).

\section{Results}\label{sec:results}

We now provide a detailed discussion of the nonlinear evolution of the TS dynamo inside a neutron-star merger remnant. The preceding evolution of this system is described in detail in \citet{Ng2024c}, and we focus only on aspects of magnetic-field evolution, AM transport, and the launching of winds from the system. We primarily describe our fiducial $\xi_{\rm TS}=1$ model but also comment on differences from other configurations.

\subsection{Dynamo activation and field amplification}\label{sec:growth}
As discussed in Section \ref{sec:methods}, the TS dynamo requires a stably stratified background, as evidenced by {real-valued} Brunt--V\"ais\"al\"a frequencies, $N_\mu$ and $N_T$, under compositional and thermal gradients, respectively.
These are shown in Fig.~\ref{fig:fig1} at an early
{stage} ($40$ ms) and midway through the evolution ($300$ ms) {for the fiducial model}. We can see that almost the entire remnant is stably stratified, consistent with earlier findings \citep{Radice2023}.
We begin by discussing the early state.
The composition contribution, $N_\mu$, is initially localized deep inside the remnant core. Referring to Eq.~\eqref{eq:N_bv},
since $\partial_{n_P}\alpha>0$ in the outward
pressure-normal direction, a stabilizing composition contribution requires
$C_{L,\mu}<0$ (see Eq.~\ref{eq:CL_appendix}).
In the dense neutron-rich core, $(\partial e/\partial Y_e)_{P,s}$ is typically positive: increasing $Y_e$ reduces the neutron-excess pressure support, so that a larger energy density is required along an isobaric and isentropic
displacement. Hence a finite $N_\mu$ requires a negative $Y_e$-gradient, i.e., $\partial_{n_P}Y_e<0$.
The $Y_e$ decreases slightly outward and is nearly flat over the core because of
trapped-neutrino equilibrium~\citep{Radice2023}, thus localizing $N_\mu$ deep in the core.
Interestingly, the high-latitude parts of the core become semi-transparent to
neutrinos while still maintaining relatively high densities. This allows a large $Y_e$ gradient to develop near the core surface, producing the thin
layers of finite $N_\mu$, which persist until the end of the simulation {($0.4$ s)}.
In contrast, the thermal contribution, $N_T$, extends over a much larger
part of the remnant. The outward entropy gradient gives a stabilizing thermal Ledoux contribution over broad regions, despite the negative temperature gradient, and its magnitude is generally larger. This component thus provides the main source of buoyancy inside the {remnant}, consistent with the findings of \citet{Radice2023}.
While matter in the disk is not stably stratified, the disk region does not contribute to the TS dynamo.

However, stable stratification alone does not lead to TS activation.
The upper-half portions of the third column of Fig.~\ref{fig:fig1} show that the TS branch is highly geometry-dependent and localized mainly in the
high-latitude parts of the core. This is consistent with the fact that unstable Tayler modes in stars are confined to an angle of $\pi/3$ from the rotation axis \citep{Ma:2019cpr}, {even though} we do not explicitly impose this constraint here. 
Initially, the neutrino-viscosity constraint allows TS activation only in fluid regions where the toroidal field strength exceeds the threshold, $B^{\rm tor} = |\psi^2 R B^{\phi}|\sim 1$--$3.5\times 10^{15}~\rm{G}$. This range is comparable to that reported by \citet{Reboul-Salze2025}, even when considering our use of a more accurate neutrino-viscosity prescription.

Most importantly, we find that these thermodynamic conditions persist throughout the simulation, although their geometric locations change as the remnant contracts, as evidenced by the bottom row of Fig.~\ref{fig:fig1} at $300$ ms. However, as we will discuss, AM transport efficiently removes the positive shear gradient, so the TS dynamo naturally saturates and {slowly switches off} at late times.

In addition to the TS dynamo, we also employ an MRI subgrid dynamo model.
The lower-half portions of the third column show the MRI-dynamo coefficient,
which is concentrated mainly in the outer negative-shear layer and disk region.
We can see that the MRI and TS dynamos are, as expected, active in complementary parts of the remnant, with the MRI-driven dynamo mostly active in the outermost layers of the neutron-star remnant and inside the disk \citep{Kiuchi2017}.
After the remnant reaches global saturation, both dynamo processes can still be activated intermittently until the shear is completely erased. However, $\kappa$ near the core appears more transient (see lower row of Fig.~\ref{fig:fig1}). This is because significant AM transport continuously readjusts the centrifugal equilibrium during contraction, thereby modifying the buoyancy and $\Omega$, which
can be seen in the lower row of Fig.~\ref{fig:fig1};
the details are discussed in later sections.


\begin{figure}
    \hspace*{-0.02\textwidth}
    \includegraphics[width=0.48\textwidth]{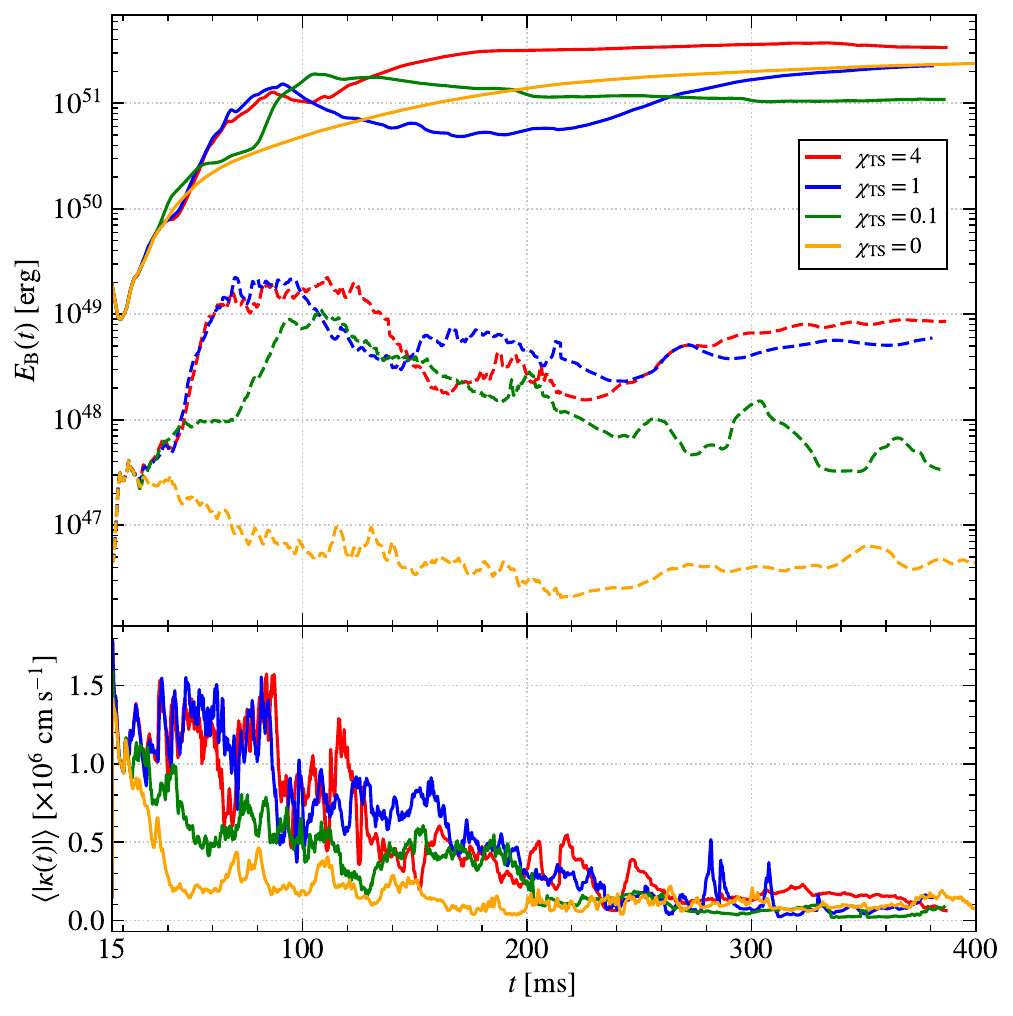}
    \caption{\textit{Upper:} Time evolution of the toroidal (solid) and poloidal (dashed) components of the magnetic energy in the merger remnant for all simulated dynamo saturation parameters, $\xi_{\rm TS}$. 
    \textit{Lower:} Time evolution of mass-averaged $|\kappa|$, which, for $\xi_{\rm TS}>0$ models, is primarily TS-dominated.
    Different colors correspond to the $\xi_{\rm TS}=0$ model, without the TS dynamo, and to models with different values of the saturation parameter $\xi_{\rm TS}$.
    }
    \label{fig:fig2}
\end{figure}
\begin{figure}
    \hspace*{-0.03\textwidth}
    \includegraphics[width=0.48\textwidth]{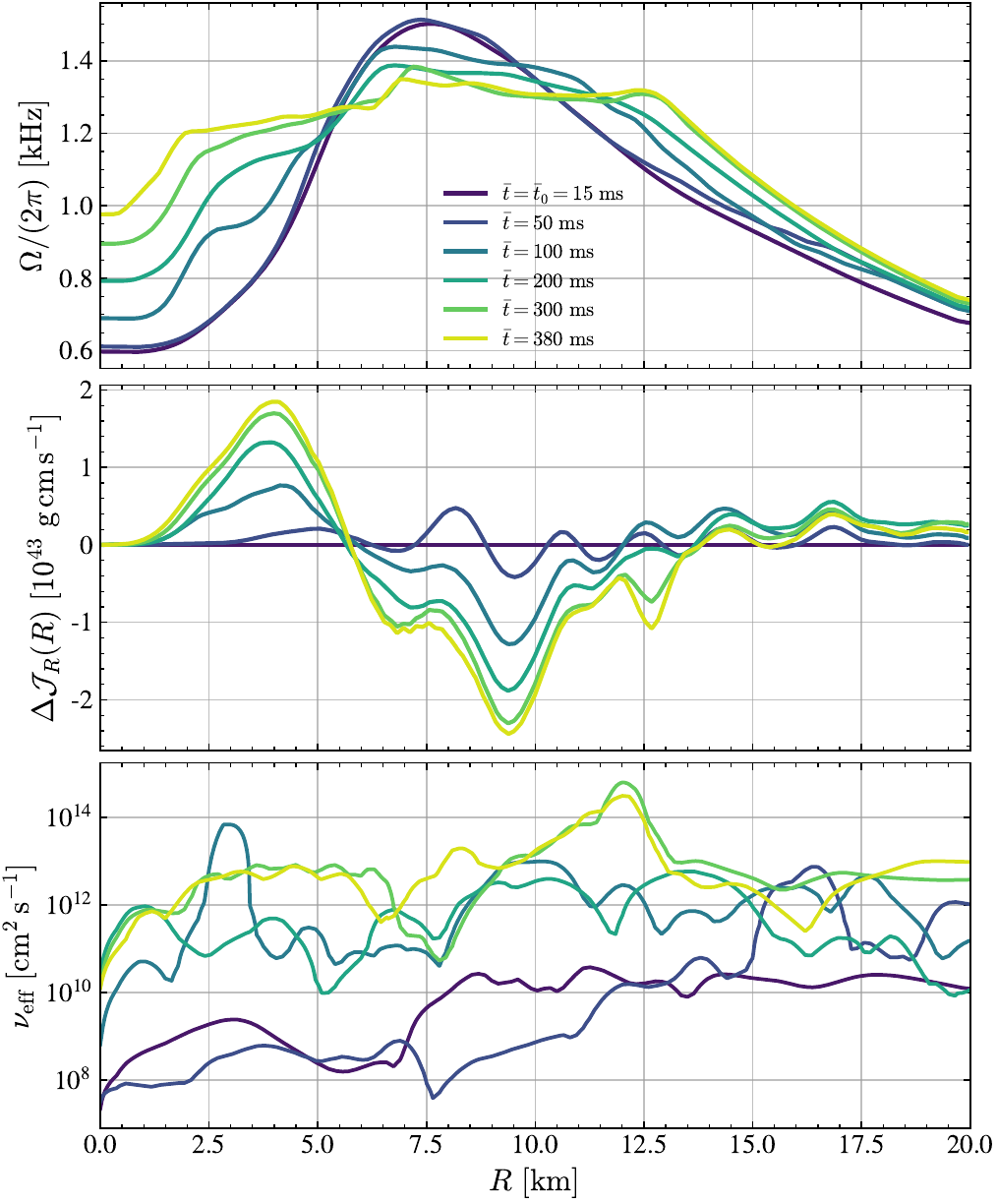}
    \caption{
    Radial profiles {of angular momentum (AM) transport} for the fiducial model {($\xi_{\rm TS}=1$)} at $z=0$.
    From top to bottom, we show the angular frequency,
    the local net AM change due to radial transport ($\Delta \mathcal{J}_{R}$ in Eq.~\eqref{eq:deltaJradial}), and the effective shear viscosity associated with the Maxwell stress $\nu_{\rm eff}$. Curves of different colors correspond to
    snapshots between $\bar t=15~{\rm ms}$ and $\bar t=380~{\rm ms}$.
    }
    \label{fig:fig3}
\end{figure}

We now discuss the general magnetic-field amplification in our simulation, which is shown in Fig. \ref{fig:fig2}. We compute the toroidal and poloidal components of the magnetic energy as $\frac{1}{2}\int {\rm d}V \sqrt{\gamma}\, B_\phi B^\phi$ and $\frac{1}{2}\int {\rm d}V \sqrt{\gamma}\, (B^2 - B_\phi B^\phi)$, respectively.
In the TS models ($\xi_{\rm TS}>0$), the subgrid dynamo leads to efficient amplification of both poloidal and toroidal magnetic fields during the first $100\, \rm ms$.  
Subsequently, the poloidal field is wound up by differential rotation, amplifying the toroidal component and leading to slow AM transport \citep{Shapiro04}. We can see that this transition occurs when the initial exponential magnetic-field
amplification stage saturates the TS dynamo criterion \eqref{eq:kappa_tar} and the subgrid model {starts to deactivate}. {The growth of 
poloidal fields slows down as well.} 
This occurs at 
{$\bar t \simeq 130~\rm{ms}$ for 
$\xi_{\rm TS} = 4$, $\bar t \simeq100~\rm{ms}$ for $\xi_{\rm TS}=1$ and $\bar t \simeq 120~\rm{ms}$ for $\xi_{\rm TS} = 0.1$.} 
{Figure \ref{fig:fig2} shows that the TS dynamo then slowly switches off over a timescale of $100\, \rm ms$ following initial saturation, with small episodes of repeated activity in local pockets of remaining differential rotation (see also Fig. \ref{fig:fig3}).}

{After} the growth phase, the mass-averaged magnetic-field strength can reach {$|B|\sim3\times10^{16}\,{\rm G}$}, while in the positive-shear core the {mass-averaged} poloidal field grows from $\sim10^{14}$ to $10^{15}\,{\rm G}$.
In the core, the magnetic field can even reach
{$B^{\rm tor}\sim5.0\times10^{17}~{\rm G}$ and
$B^{\rm pol}\sim1\times10^{16}~{\rm G}$ for the fiducial model.}\footnote{Such strong magnetic field conditions may also cause feedback onto the equation of state \citep{Most2025}.}
These core field strengths and the saturation
timescale are broadly consistent with the one-zone model of
\citet{Reboul-Salze2025}, in which an initial radial field {$B^{R}\sim10^{14}~{\rm G}$ reaches the saturation with large core fields of
$B^{\rm tor}\sim10^{17}~{\rm G}$ and
$B^{\rm pol}\sim10^{16}~{\rm G}$}. 

{Unlike in \citet{Reboul-Salze2025} {with a timescale of $20~\rm{ms}$ for the growth phase {of dynamo}}, however, {in our case,} TS dynamo remains active for much longer because pockets of differential rotation in TI-unstable regions are not immediately removed, as they are in a one-zone model.}

In contrast, the $\xi_{\rm TS}=0$ model, which uses only the MRI subgrid model and serves as a control case, shows a much weaker and shorter poloidal-amplification stage. Although the toroidal energy can still grow through magnetic winding and eventually approach a comparable global level, the poloidal component is not efficiently sustained and decays after the early phase.

The dependence on $\xi_{\rm TS}$ is moderate.
{Varying $\xi_{\rm TS}$ produces at most a factor-of-three difference in the total magnetic energy between the $\xi_{\rm TS}=4$ and $\xi_{\rm TS}=0.1$ cases. This variation is substantially smaller than the order-of-magnitude differences found for the MRI dynamo by \citet{Most2023b}, where the saturation parameter more directly sets the final magnetic-energy scale.
} In the TS branch, the saturation level is more tightly constrained by buoyancy, neutrino viscosity, and the required hierarchy of characteristic frequencies, none of which are free parameters of our model; this adds to its robustness.

\subsection{Efficient angular momentum transport}\label{sec:AM}

We proceed by discussing the resulting AM transport inside the merger remnant.
The TS dynamo taps rotational energy and momentum to amplify the fields, generating Maxwell stresses that exert torques on the
differentially rotating remnant, transporting AM from
faster-rotating regions to slower-rotating regions and thereby reducing the shear.
To quantify this process, Fig.~\ref{fig:fig3}
shows radial profiles of the fiducial model $(\xi_{\rm TS} =1)$ in the equatorial plane, $z=0$, at different times throughout its evolution.
We define the net AM change, $\Delta\mathcal{J}_{R}$, due to radial transport
as
\begin{equation}
\Delta \mathcal{J}_{R}
\equiv
-\frac{\partial}{\partial R}
\int_{t_0}^{\bar t}
dt'\,
\left[
\dot J^{\rm adv}_{R}(R,t')
+
\dot J^{\rm mag}_{R}(R,t')
\right] ,
\label{eq:deltaJradial}
\end{equation}
where $\dot J^{\rm adv}_{R}=
2\pi
\int_{\rho>10^{12}{\rm g~cm^{-3}}}
dz\,
\alpha\sqrt{\gamma}\rho h u^R u_\phi$
and
$
\dot J^{\rm mag}_{R}
=
2\pi
\int_{\rho>10^{12}{\rm g~cm^{-3}}}
dz\,
\alpha\sqrt{\gamma}
\left(-b^R b_\phi\right)$
denote the surface-integrated outward radial AM fluxes at $R$, associated with advective and magnetic transport, respectively, and $u^\mu$ is the fluid four-velocity. Positive $\Delta\mathcal{J}_{R}$ corresponds to a local net AM gain by the radial shell.
The time-integrated quantity $\Delta\mathcal{J}_R$ shows that
the radial AM redistribution is most efficient during the
first $\sim 200~{\rm ms}$, when the shear is largest and the TS dynamo is still active (see Fig. \ref{fig:fig2}).
{We can also correlate this with the secular decay in poloidal magnetic energy shown in Fig.~\ref{fig:fig2} consistent with winding and braking of the remnant \citep{Shapiro04}.
In detail, the sign of $\Delta\mathcal{J}$ indicates that AM is
removed from the region 
$R\simeq 6$--$13~{\rm km}$ and transported
inward to the core, $R\lesssim 6~{\rm km}$ {and outward to the inner disk}. Close to the rotation axis at $R=0$,
the local TS amplification is weak or suppressed;
matter can still be magnetized by advection and radial transport from neighboring layers. Since the moment of inertia is very small near the axis, a small net $\Delta\mathcal{J}$ can produce a large change in $\Omega$.

We can contrast this evolution with the change in differential rotation, as evidenced by the rotation profile $\Omega (R)$.
The $\Omega$ profile with $R<12.5~\rm{km}$  is largely smoothed by $\bar t\simeq 380~{\rm ms}$ and shows {less significant changes} afterward, forming a quasi-uniformly rotating core.
In detail, we find that the characteristic spin-down timescales for the outer core to be quasi-uniformly rotating, $\tau_{\rm flat}$ are around {$3280~\rm{ms}$, $103~\rm{ms}$ and $125~\rm{ms}$ at $\bar t = 50~\rm{ms}$, $100~\rm{ms}$ and $300~\rm{ms}$}, respectively. 
$\tau_{\rm flat}(t)$ is defined by 
\begin{equation}
\tau_{\rm flat}(t)
\simeq
\left|
\frac{
\mathcal{A}_{\Delta\mathcal{J}}(t)
-\mathcal{A}_{\Delta\mathcal{J}}(t_{\rm flat})
}{
\dot{\mathcal{A}}_{\Delta\mathcal{J}}(t)
}
\right|,
\end{equation}
where $\mathcal{A}_{\Delta\mathcal{J}}(t)
\equiv
\int_{R_a}^{R_b} dR\,\Delta\mathcal{J}_R(R,t).$
{Here, we define $\bar t_{\rm flat}=380~{\rm ms}$, $R_a=6~{\rm km}$, and
$R_b=13.5~{\rm km}$ for outer core.}
}

\begin{figure}[t]
    \hspace*{-0.03\textwidth}
    \includegraphics[width=0.48\textwidth]{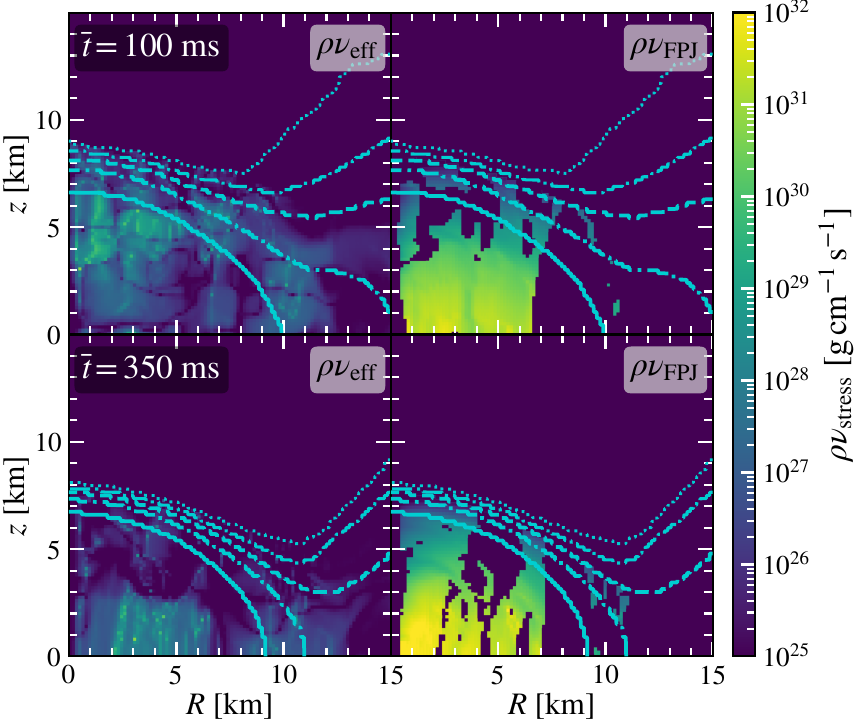}
    \caption{{{Shear} viscosity for the Tayler-Spruit dynamo at two different post-merger times, ${\bar t}$. Shown are the {\it (left)} effective viscosity {computed from the simulation with $\xi_{\rm TS}=1$}, $\nu_{\rm eff}$, {\it (right)} analytic {prescription} used in stellar evolution, $\nu_{\rm FPJ}$ \citep{Fuller2019}. Here, $\rho$ is the baryon rest-mass density.}
    }
    \label{fig:fig_nueff}
\end{figure}

We can also infer the effective shear viscosity, $\nu_{\rm eff}$, to which our runs would correspond had they been carried out in viscous hydrodynamics. This is defined as
$
{\nu_{\rm eff} =
\left|(b^R b_\phi) /
(\rho h R^2\,\partial_R\Omega)\right|}
$.
The viscosity $\nu_{\rm eff}$ measures the stress
per unit shear. It increases toward the end of the
simulation because both $B^R$ and $B^\phi$ remain strong after the
differential rotation has been substantially smoothed.
{When comparing spatial location and profiles with the analytic prescription, $\nu_{\rm FPJ}$, by \citet{Fuller2019} in Fig.~\ref{fig:fig_nueff}, 
{we find that the regions of 
occurrences are very different 
in both time moments after saturation, as 
$\nu_{\rm FPJ}$ is restricted only by the occurrence of $N^2_{\rm BV}>0$. Also, 
the strength we infer in our subgrid prescription is at least an order of magnitude smaller. This makes us caution the application of such prescriptions derived for stars for different geometries of the background field and rotation profile to the 
cases of merger remnant and proto neutron star with significant neutrino viscosity and similar $\rm{Pm}$ values}. More work also using models specifically adapted for neutron star conditions will likely be needed \citep{Skoutnev2024}.}
\begin{figure}
    \centering
    \includegraphics[width=0.48\textwidth]{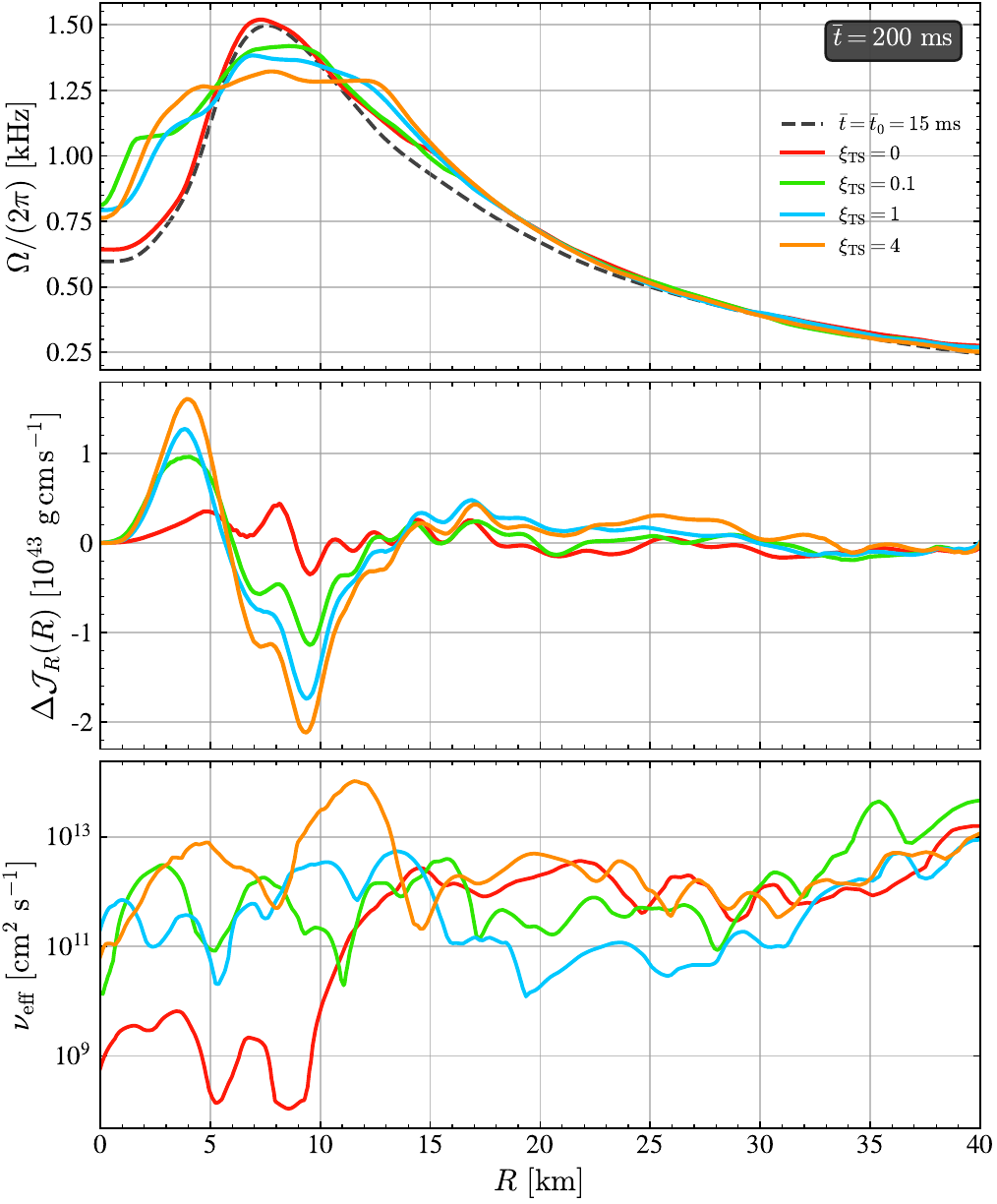}
    \caption{
    Same as Fig.~\ref{fig:fig3}, but showing
    {all cases with different dynamo saturation parameters} for an intermediate time $\bar t = 200~\rm{ms}$. Black dashed line
    corresponds to the initial angular velocity.
    }
    \label{fig:fig4}
\end{figure}
It is also informative to compare the net spatially (and temporally) varying viscosity from the TS dynamo with constant-viscosity approaches previously pursued in the literature {(e.g., \citet{Fujibayashi2017b,Fujibayashi2020b})}. In particular, we find that $\nu_{\rm eff}$ remains {a few} orders of magnitude lower than the weakest analytically prescribed shear viscosity applied to the remnant core in viscous simulations~\citep{Fujibayashi2020b}, making the realistic modeling of AM transport in the remnant a crucial ingredient for long-term postmerger evolutions.



\begin{figure}
    \centering
    \includegraphics[width=0.5\textwidth]{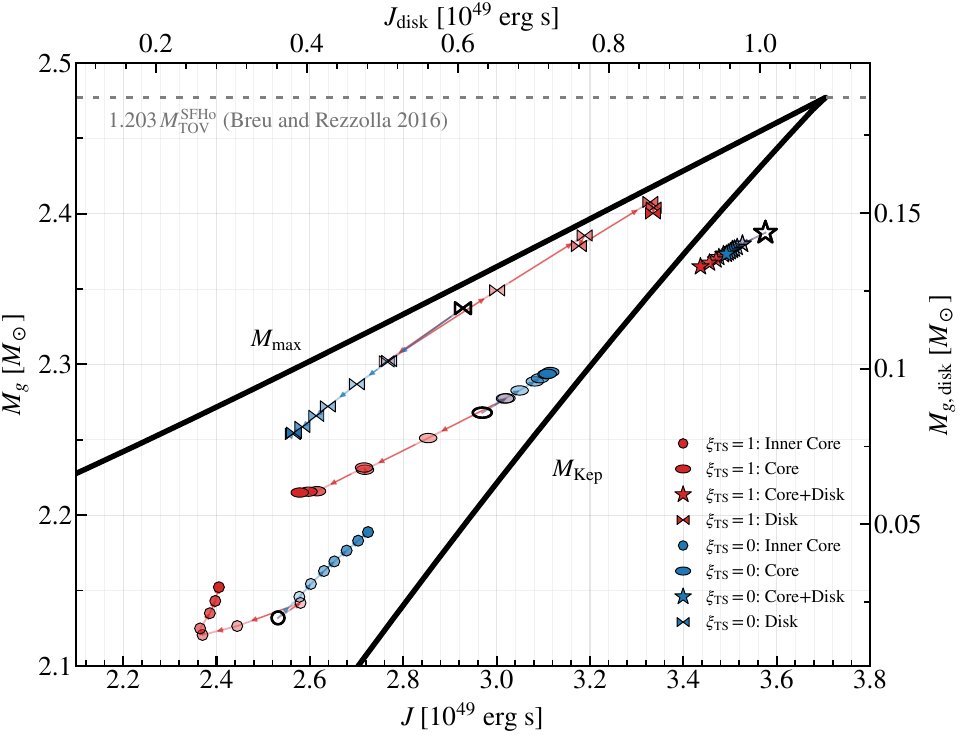}
    \caption{Evolution of gravitational mass, $M_g$, and {angular momentum (AM)} for the remnant core, disk, and total core--disk system (as indicated by the different markers). Note that the disk mass, $M_{\rm disk}$, axis is shown separately on the right. The inner core, core and disk are defined by $\rho>10^{14}$, $>10^{13}$ and $<10^{13}~{\rm g\,cm^{-3}}$, respectively, and are represented by {circle,} ellipse and bow-tie markers; stars denote the total core--disk quantities. Disk quantities are shown on the upper and right axes. Hollow markers indicate the initial state at $\bar t=15~{\rm ms}$, while filled markers show subsequent snapshots separated by approximately $50~{\rm ms}$ and connected by thin lines. Darker shades correspond to later times, and colors distinguish the models. The lower and upper solid black curves denote the mass-shedding limit, $M_{\rm Kep}(J)$, and the maximum mass supported against gravitational collapse, $M_{\rm max}(J)$, respectively, estimated using \citet{Margalit2022}, \citet{Breu2016}, and \citet{Most2020c}. The grey dashed line marks the maximum mass supported by uniform rotation at the Keplerian limit, $M_{\rm max}(J_{\rm Kep})=1.203\,M_{\rm TOV}^{\rm SFHo}$, where $M_{\rm TOV}^{\rm SFHo}=2.06\,M_\odot$~\citep{Breu2016}.
    The initial gravitational masses represented by the hollow markers are $2.13$, $2.27$, $0.12$, and $2.39~M_{\odot}$ for the inner core, core, disk, and total system, respectively.
    }
    \label{fig:fig5}
\end{figure}

\begin{figure*}
    \centering
    \includegraphics[width=1.0\textwidth]{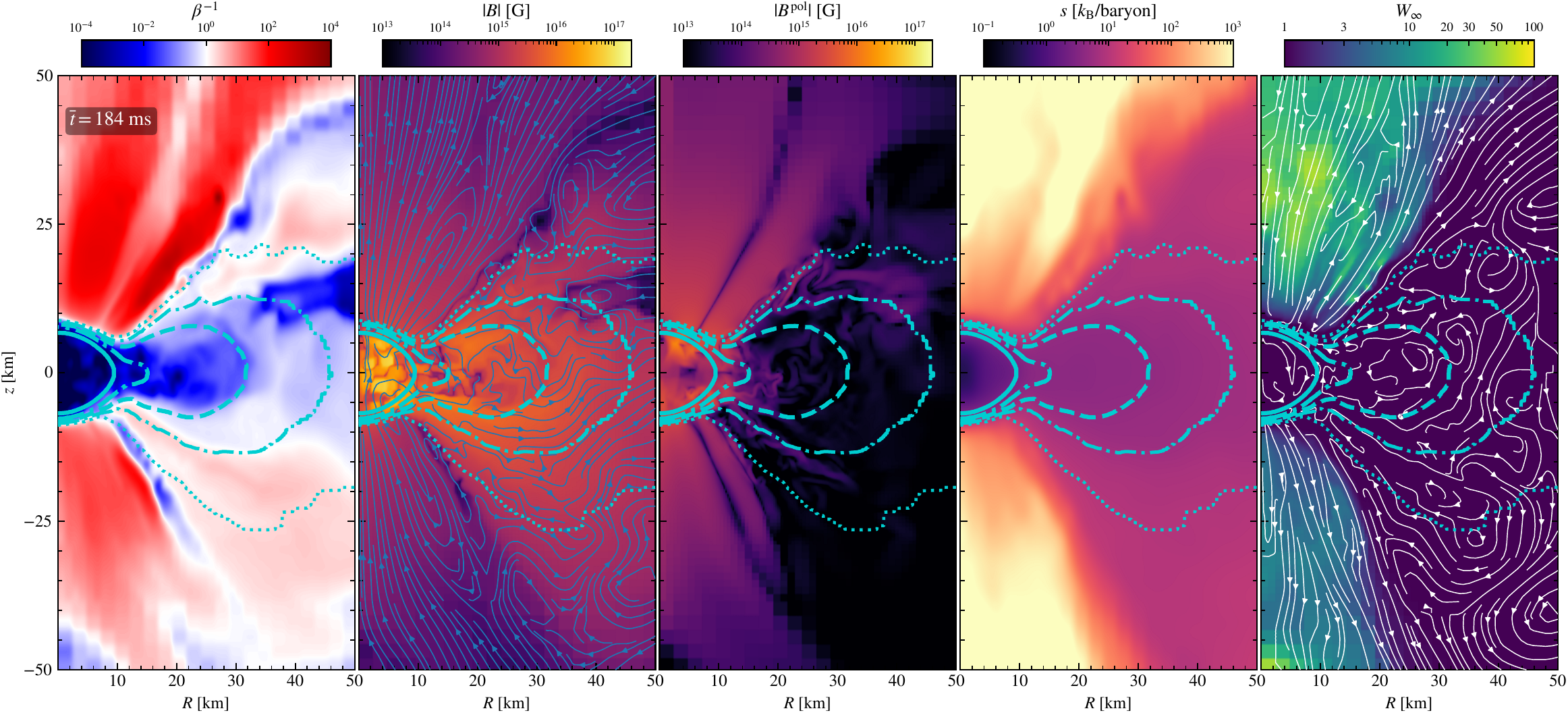}
    \caption{
    Characteristic properties of outflows from the neutron-star remnant for the fiducial $\xi_{\rm TS}=1$ case. The meridional plane is shown at time $\bar t=184~{\rm ms}$. {\it (Left to right)} Inverse plasma beta, $\beta^{-1}$, magnetic-field strength, $|B|$, poloidal-field strength, $|B^{\rm pol}| = \sqrt{B^2 - B_\phi B^\phi}$, entropy per baryon, $s$, and the asymptotic Lorentz factor, $W_\infty$, of unbound material at infinity.
    Streamlines in the $|B|$ and $W_\infty$ panels indicate the poloidal magnetic-field and velocity-field structures, respectively. Cyan contours represent densities $\rho$ of $10^{14}$, $10^{13}$, $10^{12}$, $10^{11}$, and $10^{10}\,\rm{g~cm^{-3}}$.
    }
    \label{fig:fig6}
\end{figure*}


It is now instructive to clarify how strongly or weakly AM transport depends on the parameters of our subgrid model.
Figure~\ref{fig:fig4} compares the radial structure of the remnant at
$\bar t=200~{\rm ms}$ for the models considered. The $\xi_{\rm TS}=0$
model does not exhibit any substantial AM transport inside the core and remains {differentially} rotating.
The $\xi_{\rm TS}=0$ case produces only a weak net gain/loss
of AM, whereas the TS models transport AM from the
core to $R\gtrsim20~{\rm km}$. In this outer region,
the $\xi_{\rm TS}=0$
model only mildly loses AM, while its $\nu_{\rm eff}$ becomes comparable to that of the TS models at $R\gtrsim35~{\rm km}$, {inside the disk}. This indicates that the MRI-dominated transport is recovered at large radii, while the TS dynamo provides additional Maxwell stress and positive AM transport {into the inner disk} on top of the MRI contribution. However, as we will discuss later, the AM injection into the disk region from the {core} is only effective for the TS models.
This confirms that the TS dynamo is indeed the main source of AM transport in
the core region, as probed by our simulations.

The TS models with different $\xi_{\rm TS}$ show similar AM transport behavior to the fiducial case reported in Fig.~\ref{fig:fig3}. The differences are modest because the amplification of the magnetic field is not just limited by saturation, but also by the physical constraints imposed in our model (see Fig.~\ref{fig:fig1}).
This convergence and self-limiting behavior demonstrate that our dynamical subgrid model captures two important feedback effects {that simpler one-zone models or
analytical viscosity prescriptions
do not consider~\citep{Fujibayashi2020b,Reboul-Salze2025}}.
These include an evolving rotation profile, with
significant AM redistribution depleting the shear and driving contraction, as well as the flattening of the composition and thermal gradients,
which removes or relocates the local
stratification conditions (e.g., in the deep core, see Fig.~\ref{fig:fig1}) required for the TS dynamo \eqref{eqn:TS_neutrino_cond}. {Overall, this leads to a timescale for reaching quasi-uniform rotation ($> 100~\rm{ms}$ after saturation) that is at least an order of magnitude longer than predicted by the estimation of one-zone model~\citep{Reboul-Salze2025}.}

As a general trend, increasing $\xi_{\rm TS}$ locally enhances the effective viscosity by about an order of magnitude, but since the viscosity changes over time, the resulting AM redistribution does not change by the same amount. Indeed, comparing $\Delta \mathcal{J}_{R}$ in Fig.~\ref{fig:fig4}, the net changes in local {gain of AM are within $50\%$ comparing $\xi_{\rm TS}=4$ and $\xi_{\rm TS}=0.1$}.
As secondary consequences, the transport of mass and AM affects the bulk properties of the remnant. The reduced centrifugal support in TS-dynamo cases leads to stronger contraction, a higher central density, and a slightly hotter remnant.

One important implication of AM transport in the remnant concerns its ultimate fate, especially for {higher}-mass remnants than the one considered here that can form black holes \citep{Margalit2022}, with important implications for the remnant lifetime \citep{Gill2019,Beniamini:2021tpy} and for inferring the maximum mass of neutron stars \citep{Margalit2017,Rezzolla2017,Ruiz2017,Shibata2019,Nathanail2021}.
To connect the AM transport inside the remnant to an effective sequence of rotating equilibria that models the remnant core, we extract effective trajectories in the $(J,M)$ plane separately for the inner core, core, disk, and combined system (Fig.~\ref{fig:fig5}).

We begin by focusing on the core region. As discussed in Figs. \ref{fig:fig3} and \ref{fig:fig4}, the core starts {in differential rotation}. Comparing the AM of the core alone, we find that initially it is close to the Keplerian (mass-shedding) limit, $M_{\rm Kep}$, of an equally massive, uniformly rotating configuration. As the AM transport proceeds {with $\xi_{\rm TS}=1$}, the core region efficiently spins down and moves away from the mass-shedding limit.
{Instead, because of the low system mass, the trajectory does not intersect the maximum-mass limit, $M_{\rm max}$, for uniform rotation by the end of the simulation, nor does it appear likely to do so at later times (while the disk is still present\footnote{On timescales much longer than the dynamical timescale, pulsar spin-down can trigger collapse of the remnant, e.g., \citet{Ravi2014}.}).}
We can crudely estimate the mass necessary for TS-dynamo-induced collapse based on our simulation.
In detail, we find that TS-driven spin-down corresponds to a trajectory
\begin{equation}
    M_g/M_\odot = 0.15 \left[\frac{J}{10^{49}\, \rm erg\, s}\right] + 1.82\,,
\end{equation}
where $M_g$ is the gravitational mass of the hypermassive neutron star remnant, and $J$ is it's angular momentum.
Keeping the slope and moving the trajectory upward {toward $M_{\rm max}$ and along $M_{\rm Kep}$}, we can find the point where it intersects the maximum-mass curve. This leads to a critical mass {$M^{\rm crit} \approx 2.4\, M_\odot$} for SFHo above which TS-driven spin-down would trigger collapse {within a few hundred milliseconds}. For comparison, this is just below the maximum-mass limit of {$1.203 M^{\rm SFHo}_{\rm TOV}= 2.48\,M_\odot$} \citep{Breu2016}, as well as substantially below the prompt-collapse threshold of $2.9\, M_\odot$~\citep{Koeppel2019} (see also \citet{Bauswein2013,Koelsch2021,Tootle2021,Kashyap:2021wzs}).
The estimated critical mass may not be universal across different merger
configurations, as it can depend on the initial stellar spins, the EOS and 
the modified ratio between $M_{\rm max}(J_{\rm Kep})$ and $M_{\rm
TOV}$ ( e.g.,~\cite{Breu2016,Margalit2017,Musolino2023b}).

Considering the disk, the trajectories in Fig.~\ref{fig:fig5} also show that, in addition to redistributing AM within the core (defined as $\rho>10^{13}~\rm{g~cm^{-3}}$), the TS dynamo drives a net AM flux from the core into the disk (defined as $\rho < 10^{13}~\rm{g~cm^{-3}}$). This means that both $M_{\rm disk}$ and $J_{\rm disk}$ increase over time. This contrasts with the regular evolution under effective MRI-driven disk viscosity alone, in which winds carry AM and mass out of the finite-sized disk, decreasing both $M_{\rm disk}$ and $J_{\rm disk}$, as we indeed find for the $\xi_{\rm TS} =0$ model. Because these processes operate simultaneously, substantial amounts of mass are injected from the {core} into the disk {or to be unbounded (likely around $\Delta M_{\rm disk} \simeq 0.055~M_\odot$)}. This difference in effective disk mass is comparable to, and in some cases larger than, uncertainties in the disk mass from the EOS \citep{Radice2018a}.
{At the end of our simulation, the $\xi_{\rm TS}=1$ case has {$0.07~{M}_\odot$} less core mass and $0.07~{M}_\odot$ more disk mass than the $\xi_{\rm TS}=0$ case.}

\begin{figure}
    \hspace*{-0.02\textwidth}
    \includegraphics[width=0.48\textwidth]{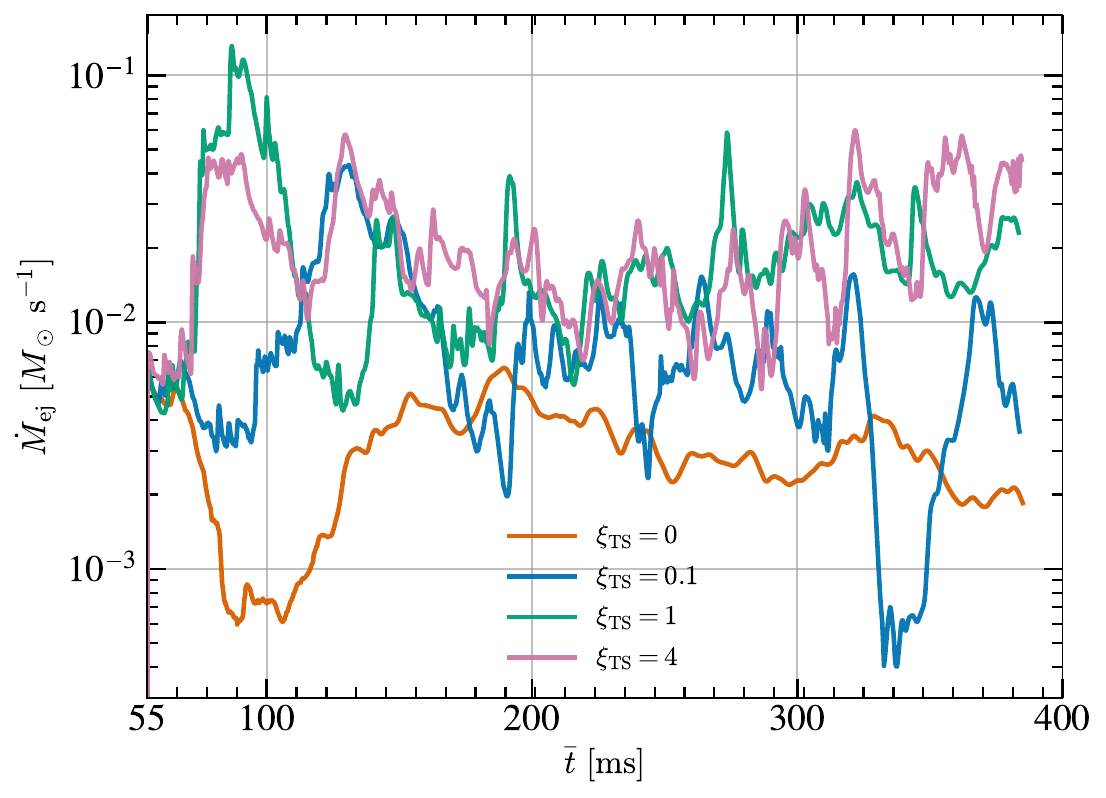}
    \caption{Mass ejection rate, $\dot{M}_{\rm ej}$, computed at a spherical radius of $550~\rm{km}$ for all simulated dynamo saturation parameters, $\xi_{\rm TS}$.}
    \label{fig:fig7a}
\end{figure}

\begin{figure*}
    \includegraphics[width=1.0\textwidth]{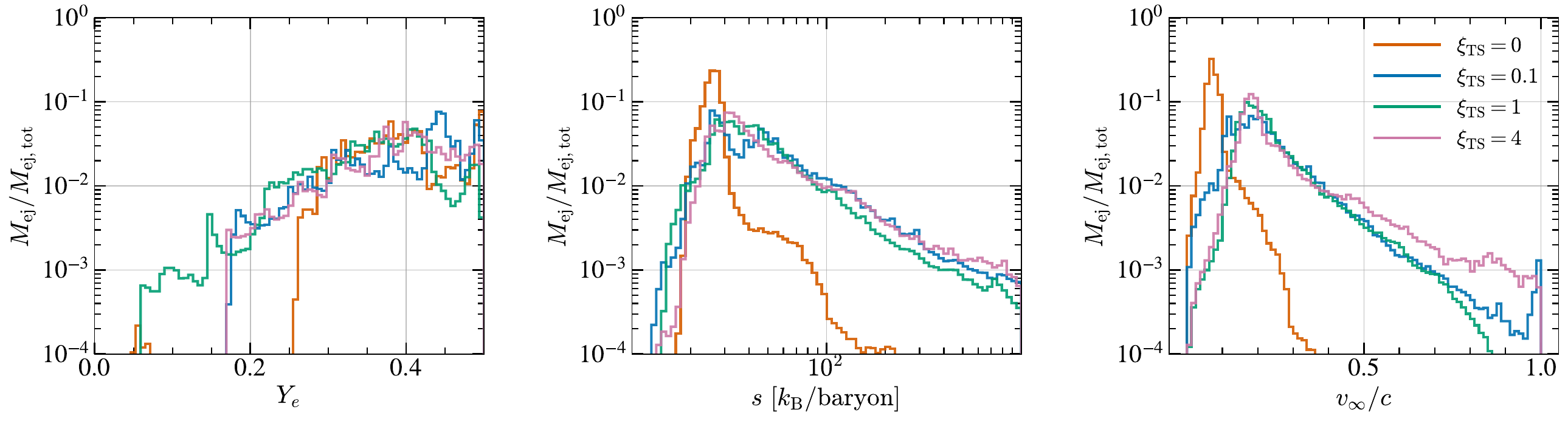}
    \caption{
    Normalized ejected mass distributions for all {dynamo saturation parameters}. {\textit{From left to right}, the panels show the distributions of electron fraction, $Y_e$, entropy, $s$, and terminal velocity, $v_\infty/c$.}}
    \label{fig:fig7b}
\end{figure*}

\subsection{Outflows}\label{sec:outflow}

Having discussed the main changes to the remnant and the disk, we now quantify how mass ejection changes when the TS dynamo is present in the system. Since we have already seen that the disk mass and AM are substantially altered, we expect large changes in the outflow, also because the injected material from the star is substantially more neutron-rich than the disk material.

\subsubsection{Magnetic tower outflows}\label{sec:jet}

We begin by discussing outflows from the remnant {for the fiducial case} (Fig.~\ref{fig:fig6}).
These are predominantly driven by magnetic-pressure-driven winds (magnetic towers) from the stellar surface (e.g., \citealt{Metzger2018}), {and can potentially power either part or all of the short gamma-ray burst \citep{Gottlieb:2023b,Lu:2015rta,Zhang:2000wx}, trough a number of dissipation mechanisms \citep{Mbarek:2026xyk}.} This picture is consistent with a number of recent GRMHD studies of merger remnants showing enhanced outflows {\citep{Moesta2020,Combi2023,Bamber2024,Wen2026}} as a result of magnetic-field amplification in the outer layers of the remnant {\citep{Most2023,Kiuchi2023}} and/or the inner region of the disk \citep{Musolino2024b}, followed by the magneto-buoyant breakout and reorganization of that field into large-scale structures{ (see also \citet{Fields2025})}.
While these studies have performed systematic investigations of this process in full 3D, naturally our axisymmetric simulations are intrinsically incomplete in this regard.


Winding in combination with the TS dynamo produces strong toroidal magnetic fields near the surface of the {merger} remnant. Consistent with previous work, these fields become Parker-unstable, rise magneto-buoyantly out of the star \citep{Kluzniak:1997nt}, and are ejected as flares that stretch the remnant field and assemble a strong, poloidally dominated funnel \citep{Most2023,Most2023b,Musolino2024b,Jiang2025} resembling a magnetic tower \citep{Lynden-Bell1996}.
This tower can then drive mass outflows {\cite{Shibata2021c}}, which we observe to have plasma $\beta^{-1}\sim10^2$--$10^4$, plasma $\sigma\sim10^2$--$10^3$, specific entropy $s\sim10^3\,k_{\rm B}/{\rm baryon}$, and Lorentz factor at infinity
$W_\infty\sim30$--$70$ (see \citet{Bekenstein1978} for the Bernoulli criterion in axisymmetric GRMHD), corresponding to an instantaneous Lorentz factor
$W\sim1.2$--$1.4$. In the fiducial case, the relativistic outflow launches from $\bar t=45~{\rm ms}$ and is sustained continuously thereafter.
Thus, the Poynting luminosity reaches $10^{50}$--$10^{51}~\rm{erg~s^{-1}}$, accounting for the slight decrease in {magnetic} energy during the saturation stage
in Fig.~\ref{fig:fig2}. By contrast, the $\xi_{\rm TS}=0$ model does not form a comparably strong
large-scale poloidal field or a highly magnetized relativistic funnel in our simulations. However, we caution that this contrasts with 3D calculations {\citep{Most2023b,Kiuchi2023,Musolino2024b}}, so a full assessment of the role of tower-driven outflows potentially enhanced by the TS dynamo must be performed in future work.

\subsubsection{Mass ejection}\label{sec:ejecta}
We now examine how these effects modify the amount and properties of the ejecta, specifically from the disk.
We extract ejected material at a spherical coordinate radius $r_{\rm ext}=550~\rm{km}$.
 The detector surface is sampled using 512 uniformly spaced polar-angle
 sections, onto which the fluid quantities are interpolated every $0.1$ ms.
Fluid elements are classified as ejecta when they satisfy the
Bernoulli criterion, $-h u_t > h_{\min}$, where $h_{\min}$ is the minimum specific enthalpy from the EOS \citep{Bovard2017}.
The distributions of the ejecta properties are weighted by the rest mass crossing each angular section
during each sampling interval and normalized by the total retained ejecta
mass. We exclude the ejecta from the first $40~\rm{ms}$ because of initial perturbations from the {$\phi$-averaging and data-transition} procedure.

We begin by discussing the overall mass ejection rate, $\dot{M}_{\rm ej}$.
Figure~\ref{fig:fig7a}
shows that the mass-ejection rate in simulations using the TS dynamo is enhanced by an order of magnitude
compared to the $\xi_{\rm TS}=0$ model. In
particular, the $\xi_{\rm TS}=1$ model continues to exhibit strong ejection
episodes several hundred milliseconds after the remnant reaches global saturation.
The accumulated ejecta masses at $\bar t\simeq400$ ms are approximately
{$M_{\rm ej,tot}\simeq1.01\times10^{-3}\,M_\odot$
($\simeq0.037\%$ of the common initial rest mass
$M_{\rm rest,0}=2.724\,M_\odot$) for $\xi_{\rm TS}=0$,
$2.89\times10^{-3}\,M_\odot$ ($\simeq0.106\%$) for
$\xi_{\rm TS}=0.1$, 
$6.86\times10^{-3}\,M_\odot$ ($\simeq0.252\%$) for
$\xi_{\rm TS}=1$, and 
$7.18\times10^{-3}\,M_\odot$ ($\simeq0.263\%$) for
$\xi_{\rm TS}=4$.
}
The fiducial case has almost three times as much ejecta as the $\xi_{\rm TS}=0$ case.

\begin{figure}
    \hspace*{-0.0\textwidth}
    \includegraphics[width=0.45\textwidth]{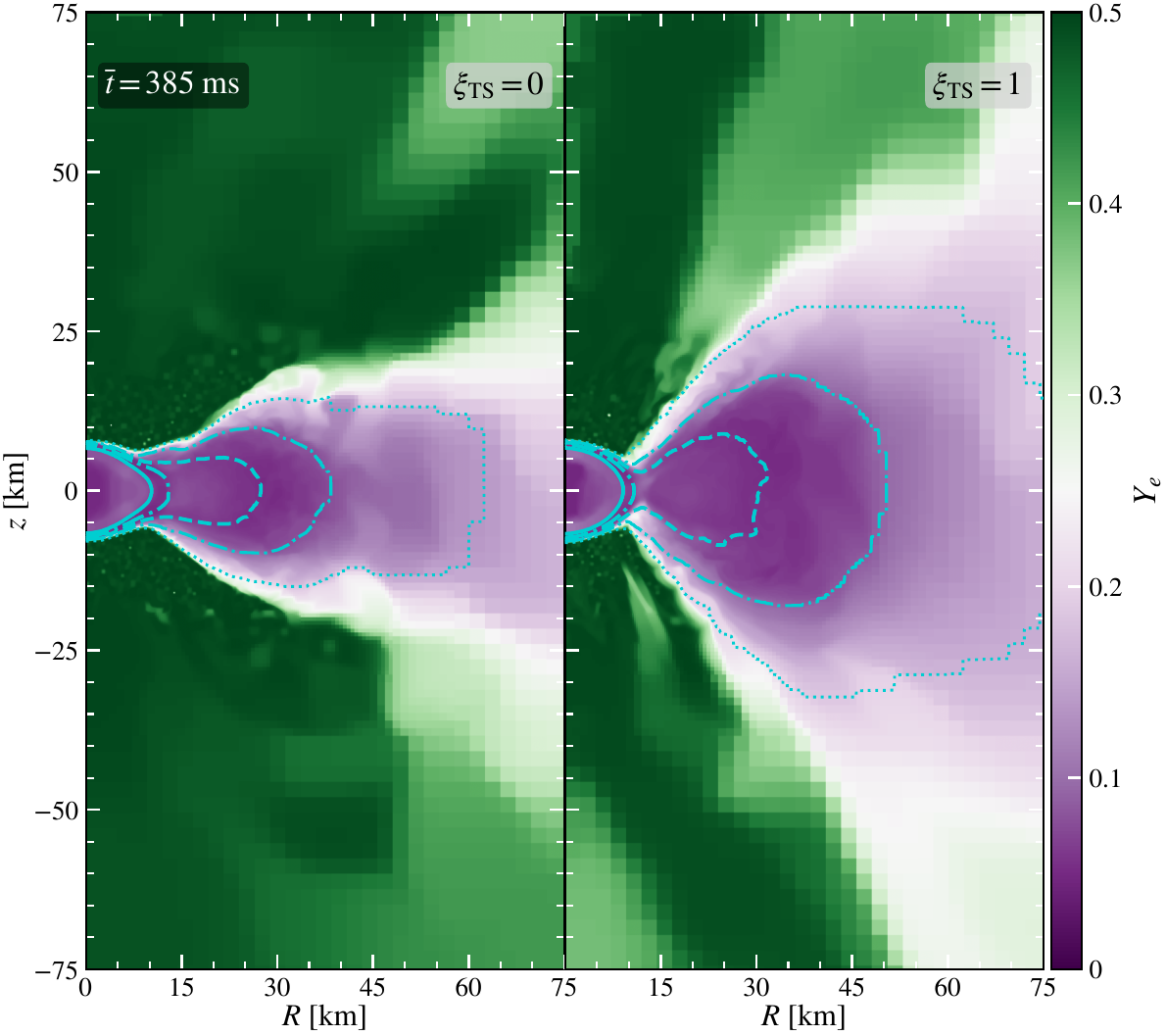}
    \caption{Electron-fraction profiles at $\bar{t}=385~\rm{ms}$ for the
    $\xi_{\rm TS}=0$ (left) and $\xi_{\rm TS}=1$ (right) models. Cyan contours denote rest-mass density as defined in Fig.~\ref{fig:fig6}.}
    \label{fig:fig8}
\end{figure}

For the TS models, most of the ejecta mass is distributed over
$50^\circ\lesssim\theta\lesssim140^\circ$, indicating that the dominant
component originates from the disk and its high-latitude atmosphere. A
smaller-mass but significantly faster component is launched through the
polar funnel (jet-like outflow in
Sec.~\ref{sec:jet}). These
high-latitude-to-polar outflows contribute part of the
high-entropy and relativistic ejecta shown in the histograms of Fig.~\ref{fig:fig7b}
in both TS-dynamo cases.
However, the substantially larger ejecta mass, broader angular distribution, higher $v_{\infty}$ and entropy, and more neutron-rich composition primarily arise from the more diffuse and thicker disk produced by the TS dynamo (Fig. \ref{fig:fig8}).

As discussed and shown in Fig.~\ref{fig:fig5}, the TS dynamo redistributes AM
and mass from the outer core to the disk (with the disk migrating to higher $M_{g,\rm disk}$ and $J_{\rm disk}$).
The readjustment of centrifugal support {and the outward transport of mass} expand the disk toward larger cylindrical radii and vertical heights (Fig. \ref{fig:fig8}).
{At $\bar{t}=385~\mathrm{ms}$, the $\rho=10^{11}$ and
$10^{10}~\mathrm{g\,cm^{-3}}$ contours in the $\xi_{\rm TS}=1$ model
extend farther in $(R,|z|)$ by approximately $(12,8)~\mathrm{km}$ and
$(20,18)~\mathrm{km}$, respectively, than in the $\xi_{\rm TS}=0$ model
(see Fig.~\ref{fig:fig8}).}
{In the $\xi_{\rm TS}=0$ case, the moment-based radiation transport produces overestimated high-$Y_e$ matter surrounding the smaller disk, consistent with the results of \citet{Zappa2023}.}
{For comparison, in the $\xi_{\rm TS}=1$ case}, neutron-rich matter from the outer core is also injected into the disk and quickly
attains $\beta$-equilibrium through weak interactions, forming a more neutron-rich disk than in the $\xi_{\rm TS}=0$ case (see Fig.~\ref{fig:fig8}). The combined effects in the TS models transport matter outward to form a larger,
more weakly bound, magnetized, and centrifugally supported disk, thus producing
a larger amount of neutron-rich, fast, and energetic ejecta.
This could enhance the postmerger ejecta component of red kilonovae from massive remnants, {potentially affecting the conventional picture of predominantly neutron-poor outflows \citep{Metzger2014}. Future work will be required to systematically investigate the impact of the TS dynamo on ejecta composition.}

\section{Conclusions}\label{sec:discussion}

We have presented the first global general-relativistic neutrino-radiation magnetohydrodynamics simulations of a binary neutron-star merger remnant incorporating a subgrid mean-field prescription for the unresolved Tayler--Spruit (TS) dynamo. Motivated by recent theoretical work suggesting that the TS dynamo may operate in the stably stratified, positive-shear cores of merger remnants \citep{Margalit2022,Skoutnev2024,Reboul-Salze2025}, we developed a relativistic dynamo closure \citep{Bucciantini2012a} that incorporates buoyancy, neutrino-viscous suppression of the Tayler instability, and local nonlinear saturation following our earlier approach \citep{Most2023b}. Combined with an MRI-driven dynamo operating in the negative-shear outer remnant and accretion disk \citep{Kiuchi2023}, this provides a unified framework for modeling unresolved magnetic-field amplification throughout the postmerger system, which cannot be captured at $\rm{Pm}\simeq 1$ (or in simulations with numerical dissipation~\citep{Skoutnev2024}).

Despite being active only in a localized region of the remnant, the inclusion of the TS dynamo has drastic consequences for the long-term evolution of the remnant. The dynamo efficiently generates Maxwell stresses that redistribute AM throughout the {remnant} core. This transport progressively removes differential rotation, drives the remnant toward quasi-uniform rotation on secular timescales, and transfers both mass and AM from the outer core into the surrounding accretion disk. As a consequence, the disk becomes more massive, more extended, and more strongly magnetized than in models without TS-dynamo action, providing a substantially larger reservoir for late-time outflows.

These changes naturally modify the observable evolution of the system. The amplified magnetic field launches a magnetically dominated polar outflow, while the larger and more weakly bound disk produces substantially more neutron-rich ejecta{, which could potentially drive a stronger red postmerger component of a kilonova.}
For remnants closer to their stability limit, this additional transport may shorten the remnant lifetime and alter the threshold for delayed gravitational collapse. Based on our simulation for the SFHo EOS~\citep{Steiner2013}
{with an initial core mass of $2.27~M_{\odot}$}, we estimate that cores with initial masses of $2.42~M_{\odot}$ will collapse {within a few hundred milliseconds}.

While our simulations highlight the importance of modeling the TS dynamo in postmerger evolutions, several caveats remain. The nonlinear saturation of the Tayler instability under neutron-star merger conditions is not yet understood from first principles \citep{Skoutnev2024}, and better models are needed to accurately characterize the timescale, the instability regions inside the remnant, the nonlinear geometric dependence, and the polar TI-driven branch~\citep{Barrere2026,Barrere2026b}.
{Additionally, trapped neutrino could contribute to the buoyancy calculation, and} neutrino-mediated thermal diffusion may weaken the effective thermal restoring force
at short radial wavelengths, enlarging the range of TI-unstable scales~\citep{Spruit2002,Fuller2019}.
Ultimately, establishing self-consistent simulations that capture the TS dynamo in merger remnants will require {3D} dissipative GRMHD simulations~\citep{Most2021d} coupled to Boltzmann transport~\citep{Foucart2018,Offermans2026_boltz} that provides physical neutrino viscosity and diffusivity.
{Finally, the approach to the TS dynamo and its impact on post-merger remnant evolution might not be unique to neutron stars, as the merger of white dwarfs has also been argued to provide conditions susceptible to the TS dynamo \citep{Shen:2011ey,Schwab:2012hd,Pakmor:2024efc}. }
\section*{Acknowledgments}
The authors are grateful to James Beattie, Andrei Beloborodov, Pavan Chawhan, Patrick Cheong, Matthew Duez, Francois Foucart, 
James Fuller, Brian Metzger, Alexander Philippov, Eliot Quataert, Luciano Rezzolla, 
Valentin Skoutnev, 
and Ellen Zweibel for insightful discussions. H.H.Y.N. is supported by a Croucher Fellowship from the Croucher Foundation. ERM acknowledges support by the U.S. National Science Foundation under Grant Nos. PHY-2541792, PHY-2309210, and AST2510568. ERM is also supported by a Research Fellowship from the Sloan Foundation and a William H. Hurt Scholarship at the California Institute of Technology.
The simulations were performed on the NSF Frontera supercomputer under grant AST21006. Additional simulations were performed on Delta at the National Center for Supercomputing Applications (NCSA) through allocation PHY210074 from the Advanced Cyberinfrastructure Coordination Ecosystem: Services \& Support (ACCESS) program, which is supported by National Science Foundation grants \#2138259, \#2138286, \#2138307, \#2137603, and \#2138296. Support also comes from the Resnick High Performance Computing Center, a facility supported by the Resnick Sustainability Institute at the California Institute of Technology.

\software{
     \texttt{EinsteinToolkit}~\citep{Loffler:2011ay},
     \texttt{Frankfurt/IllinoisGRMHD}~\citep{Etienne2015,Most2019b,Musolino2023},
     \GMUNU~\citep{Cheong2021,Cheong2023},
     \WH~\citep{Ng2024a},
	  matplotlib \citep{Hunter:2007},
	  numpy \citep{harris2020array},
	  scipy \citep{2020SciPy-NMeth}
}
\appendix
\section{Evaluation of the Brunt--V\"ais\"al\"a frequency}
\label{sec:BVfreq}
In Sec.~\ref{sec:TSI}, to avoid a geometry-induced mismatch caused by the nonspherical structure of the rapidly rotating and approximately cylindrical remnant, we evaluate $N^2_{\rm BV}$ using the adiabatic Ledoux criterion along the local outward pressure-normal direction, $n_P$, rather than along the spherical-radial direction, $r$. We define
\begin{equation}
    \hat{\boldsymbol{n}}_{P}
    \equiv
    -\frac{\boldsymbol{\nabla}P}
    {\lvert\boldsymbol{\nabla}P\rvert},
    \qquad
    \partial_{n_P}
    \equiv
    \hat{\boldsymbol{n}}_{P}\boldsymbol{\cdot}\boldsymbol{\nabla},
\end{equation}
where $\boldsymbol{\nabla}Q=\partial_R Q\,\hat{\boldsymbol{e}}_R+\partial_z Q\,\hat{\boldsymbol{e}}_z$. With this convention, $\hat{\boldsymbol{n}}_{P}$ points toward decreasing pressure, giving $\partial_{n_P}P<0$. Along the same outward pressure-normal direction, the lapse typically increases away from the compact remnant core, giving $\partial_{n_P}\alpha>0$.
In most cells, the magnitude {and sign of $N^2_{\rm BV}$} computed with the spherical-radial projection is similar to that obtained along the pressure-normal direction. However, in high-latitude regions, the spherical-radial projection, $\partial_r$, can slightly under-identify stably stratified cells with {$N_{\rm BV}^2>0$}, an effect that becomes more pronounced in lower-resolution setups. The Ledoux discriminant, $C_{\rm L}$, entering Eq.~\eqref{eq:N_bv} is defined as
\begin{equation}
\begin{split}
C_{\rm L}
\equiv{}&
\partial_{n_P} e
-
\left(\frac{\partial e}{\partial P}\right)_{s,Y_e}
\partial_{n_P} P
\\
={}&
\left(\frac{\partial e}{\partial s}\right)_{P,Y_e}
\partial_{n_P} s
+
\left(\frac{\partial e}{\partial Y_e}\right)_{P,s}
\partial_{n_P} Y_e
\\
={}&
C_{{\rm L},T}+C_{{\rm L},\mu},
\end{split}
\label{eq:CL_appendix}
\end{equation}
where $e=\rho(1+\epsilon)$ is the internal energy density, $s$ is the specific entropy,
and $C_{\rm{L}}$ is decomposed into the thermal and compositional
parts for computing $N_T$ and $N_\mu$, respectively.
We evaluate the thermodynamic derivatives using centered finite differences in the high-density nuclear EOS and store them {before the evolution.}
The derivatives in Eq.~\eqref{eq:CL_appendix} are
\begin{equation}
    \left(
    \frac{\partial e}{\partial s}
    \right)_{P,Y_e}
    =
    \frac{
    e_TP_\rho-e_\rho P_T
    }{
    \mathcal{D}_{\rm L}
    },
    \label{eq:es_appendix}
\end{equation}
and
\begin{equation}
\begin{aligned}
    \left(
    \frac{\partial e}{\partial Y_e}
    \right)_{P,s}
    =&\,
    e_{Y_e}
    +
    e_\rho
    \frac{
    P_Ts_{Y_e}-P_{Y_e}s_T
    }{
    \mathcal{D}_{\rm L}
    }
    \\
    &+
    e_T
    \frac{
    P_{Y_e}s_\rho-P_\rho s_{Y_e}
    }{
    \mathcal{D}_{\rm L}
    },
\end{aligned}
\label{eq:eYe_appendix}
\end{equation}
where $
    Q_\rho \equiv
    \left(
    \frac{\partial Q}{\partial\rho}
    \right)_{T,Y_e}$
    $,
    Q_T
    \equiv
    \left(
    \frac{\partial Q}{\partial T}
    \right)_{\rho,Y_e}
    $,
    $
    Q_{Y_e}
    \equiv
    \left(
    \frac{\partial Q}{\partial Y_e}
    \right)_{\rho,T}
    $,
    and $\mathcal{D}_{\rm L} \equiv P_\rho s_T-P_Ts_\rho$.

\section{Low-density Helmholtz-like EOS and Neutrino microphysics}
\label{app:micro}
The CompOSE\footnote{\url{https://compose.obspm.fr}} table used in this work only extends down to $T_{\rm min}\simeq0.1~{\rm MeV}$ and $\rho_{\rm min}\sim10^3~{\rm g~cm^{-3}}$, and assumes nuclear statistical equilibrium (NSE). However, low-density ejecta with $T\lesssim 0.5~{\rm MeV}$ are not expected to remain in NSE. We therefore employ a simplified and efficient analytic Helmholtz-type EOS, avoiding precomputed Helmholtz tables (e.g., \citet{Timmes2000}) to reduce memory cost.
The transition to the Helmholtz branch is used either outside the nuclear-EOS table, i.e., when $\rho<\rho_{\rm min}$ or $T<T_{\rm min}$, or in the non-NSE ejecta regime, defined by $\rho<\rho_{\rm thr}=5\times10^3~{\rm g~cm^{-3}}$ and $T<T_{\rm thr}=0.5~{\rm MeV}$.

The analytic Helmholtz EOS contains the contributions of
classical ions, a non-degenerate relativistic electron--positron gas, and photons, the latter of which can become important in hot, dilute regions.
For the ion component, the mean atomic mass $\bar A$ is taken from the low-density tail of the nuclear EOS at the transition point and supplied as a fixed input to the analytic Helmholtz branch. This provides an approximate and reasonable thermodynamic closure in the absence of realistic non-NSE nucleosynthesis output.

For the electron--positron component, we adopt the classical, non-degenerate limit of a relativistic ideal Fermi gas. The pair number density and thermodynamic contributions are evaluated using the standard modified-Bessel-function expressions, corresponding to the weak-degeneracy expansion of \citet{Blinnikov96}, thereby avoiding heavy numerical integration of the Fermi integrals.

A practical issue in extending a tabulated nuclear EOS with an analytic Helmholtz branch is the mismatch between the negative binding-energy offset included in the nuclear specific internal energy and the different zero point used by the analytic Helmholtz specific internal energy. As pointed out by \citet{Hayashi2021a}, if this mismatch is left untreated, primitive recovery may converge to unphysical states or fail. We therefore require $\epsilon$ to be continuous across the nuclear--Helmholtz interface. For our analytic Helmholtz EOS, we define the bounded matching state as $\rho^\ast=\max(\rho,\rho_{\rm thr})$ and $T^\ast=\max(T,T_{\rm thr})$, and shift the Helmholtz specific internal energy as $\epsilon_{\rm H}(\rho,T,Y_e)=\epsilon_{\rm H}^{0}(\rho,T,Y_e)+\Delta\epsilon(\rho^\ast,T^\ast,Y_e)$, where $\Delta\epsilon(\rho^\ast,T^\ast,Y_e)=\epsilon_{\rm nuc}(\rho^\ast,T^\ast,Y_e)-\epsilon_{\rm H}^{0}(\rho^\ast,T^\ast,Y_e)$. This gives $\epsilon_{\rm H}(\rho^\ast,T^\ast,Y_e)=\epsilon_{\rm nuc}(\rho^\ast,T^\ast,Y_e)$, removing the inconsistencies in $\epsilon$ between the two EOSs.
\begin{figure}
    \centering
    \includegraphics[width=\linewidth]{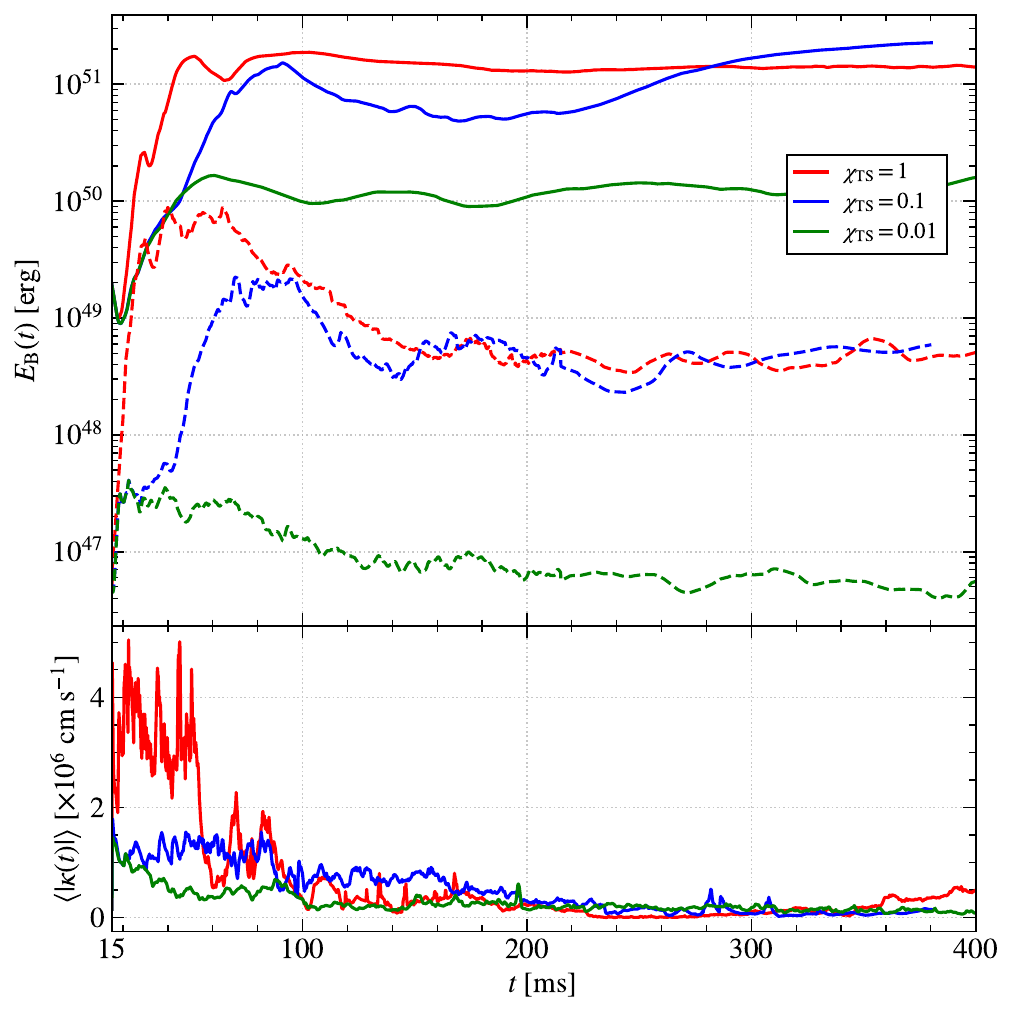}
    \caption{Same as Fig. \ref{fig:fig2}, but comparing different values of $\chi_{\rm TS}$.}
    \label{fig:chi_comp}
\end{figure}

For neutrino microphysics, we follow the same interaction set, treatments, and corrections for each interaction as in the three-species case of \citet{Ng2024c}, except for two modifications. First, plasmon decay (g) and muonic processes (c) and (d) are ignored. Second, pair processes, i.e., processes (e) and (f) therein, employ the approximate-emissivity approach~\citep{Ng2024a} instead of the kernel approach, which depends on the energy distributions/moments of neutrinos and antineutrinos, i.e., Eq.~(50) of \citet{Ng2024a}.

\section{Model uncertainties}\label{sec:uncertainty}
\subsection{Different dynamo efficiency parameter}\label{sec:diff_chi}

One major uncertainty in the model is the precise choice of $\kappa_{\rm TS}$, given different models for the Tayler-Spruit dynamo \citep{Spruit2002,Fuller2019,Skoutnev2024,Gusakov:2026bhd}. While the model by \citet{Fuller2019} we adopt present an upper bound to the angular momentum transport, the original model by \citet{Spruit2002} could be interpreted as a natural lower bound. For that model Eq. \eqref{eq:TS_fuller_fluctuations} is to be replaced with \citep{Spruit2002},
\begin{equation}
\frac{\delta B_\perp}{\bar B_{\hat\phi}}
\sim
1,
\qquad
{\delta v_\perp
\sim
\frac{\omega_{A, \rm sat}}{|\Omega|}\delta v_A,
\qquad
\delta v_A
\sim
R \omega_{A, \rm sat}}.
\label{eq:TS_spruit_fluctuations}
\end{equation}
One then arrives at
\begin{align}
    \kappa_{\rm Spruit} =  \left(\frac{q |\Omega|}{N_{\rm BV}}\right)^{4/3}\kappa_{\rm TS}\,,
    \label{eq:kappa_spruit}
\end{align}
\begin{figure}
    \centering
    \includegraphics[width=\linewidth]{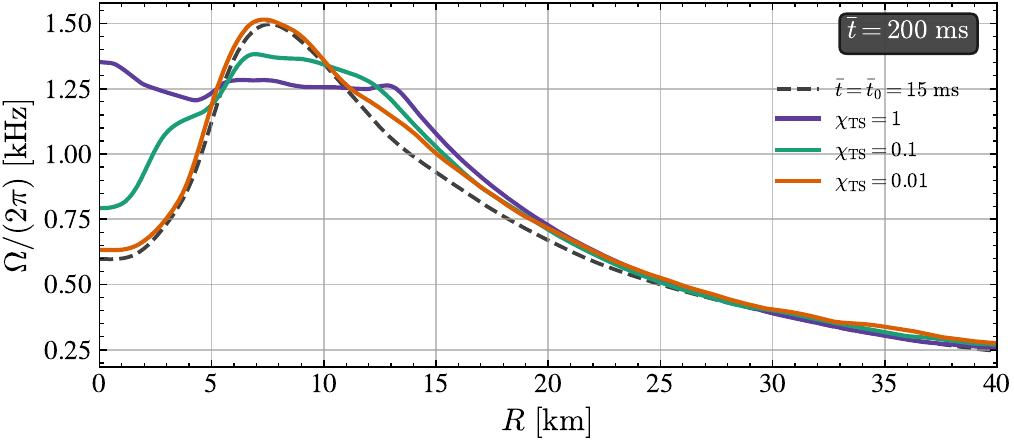}
    \caption{Same as Fig. \ref{fig:fig3}, but comparing different values of $\chi_{\rm TS}$.}
    \label{fig:omega_chi_comp}
\end{figure}

Given the characteristic ordering of the TS dynamo, this motivates
$0<\chi_{\rm TS}<1$. For representative conditions in the TS-active
core, $q\simeq0.3$, $|\Omega|\simeq6\times10^{3}\ {\rm s^{-1}}$, and
$N_{\rm BV}\simeq10^{4}\ {\rm s^{-1}}$, the original Spruit prescription gives
\[
\chi_{\rm TS}\sim
\left(\frac{q|\Omega|}{N_{\rm BV}}\right)^{4/3}
\simeq 0.1, 
\]
which matches the $\chi_{\rm TS}$ employed in simulations in main text.
To parametrically assess the uncertainty in the dynamo efficiency, we
therefore perform additional simulations with
$\chi_{\rm TS}=[0.01,\,0.1,\,1]$, spanning a strongly suppressed case,
a representative Spruit-like efficiency, and the unreduced Fuller
normalization, respectively.


Fig. \ref{fig:chi_comp} shows the evolution of magnetic energy for the
different cases. We can see that saturation is comparable for the $\chi_{\rm
TS} =0.1$ and $\chi_{\rm TS} =1$ cases, indicating comparable Maxwell stresses.
They mainly differ in the initial saturation time, which is substantially below
$100\, \rm ms$ for the $\chi_{\rm TS} =1$ case. No substantial TS dynamo action
is observed for $\chi_{\rm TS} =0.01$. Comparing the amount of angular momentum
transfered via the effective rotation curve at a fixed time (see Fig.
\ref{fig:omega_chi_comp}), we can see that both high-value cases lead to an
establishment of near uniform rotation, with a substantial spin-up of the disk
part. Therefore, we think that our results are -- within intrinsic
uncertainties about the analytic modeling of TS dynamo -- appropriately
capturing the impact of TS-driven AM transport in the post-merger remnant.
\begin{figure}
    \includegraphics[width=\columnwidth]{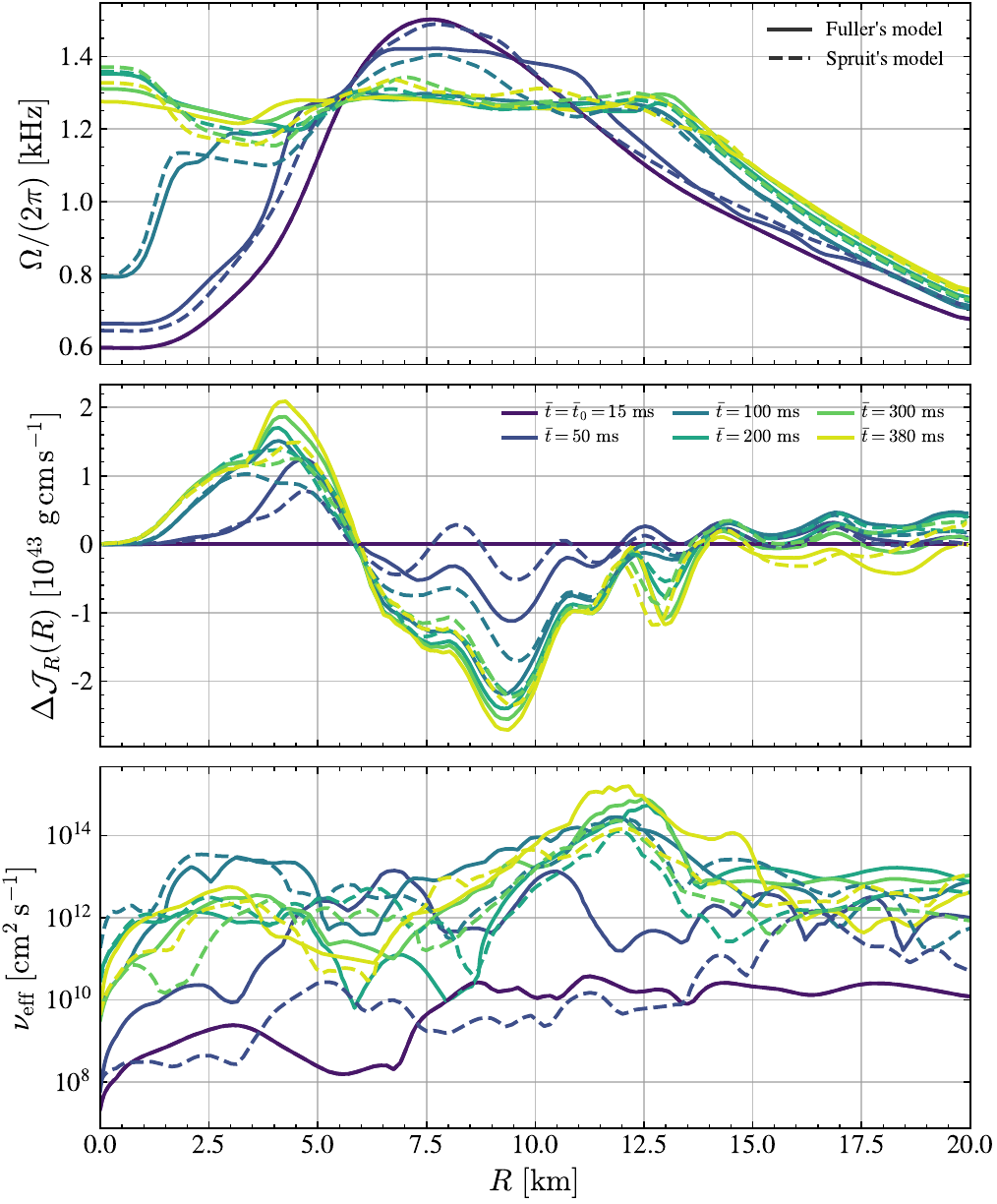}
    \caption{
    Same as Fig.~\ref{fig:fig3}, but comparing effects of AM transport at different time momemtns between
    Spruit model~\citep{Spruit2002} (dashed lines)
    and Fuller's model (solid lines) used in main text.
    We assume $\xi_{\rm TS} = 1$ and $\chi_{\rm TS} = 1$ in both cases.
    }
    \label{fig:spruit_am}
\end{figure}

\begin{figure}
    \includegraphics[width=\columnwidth]{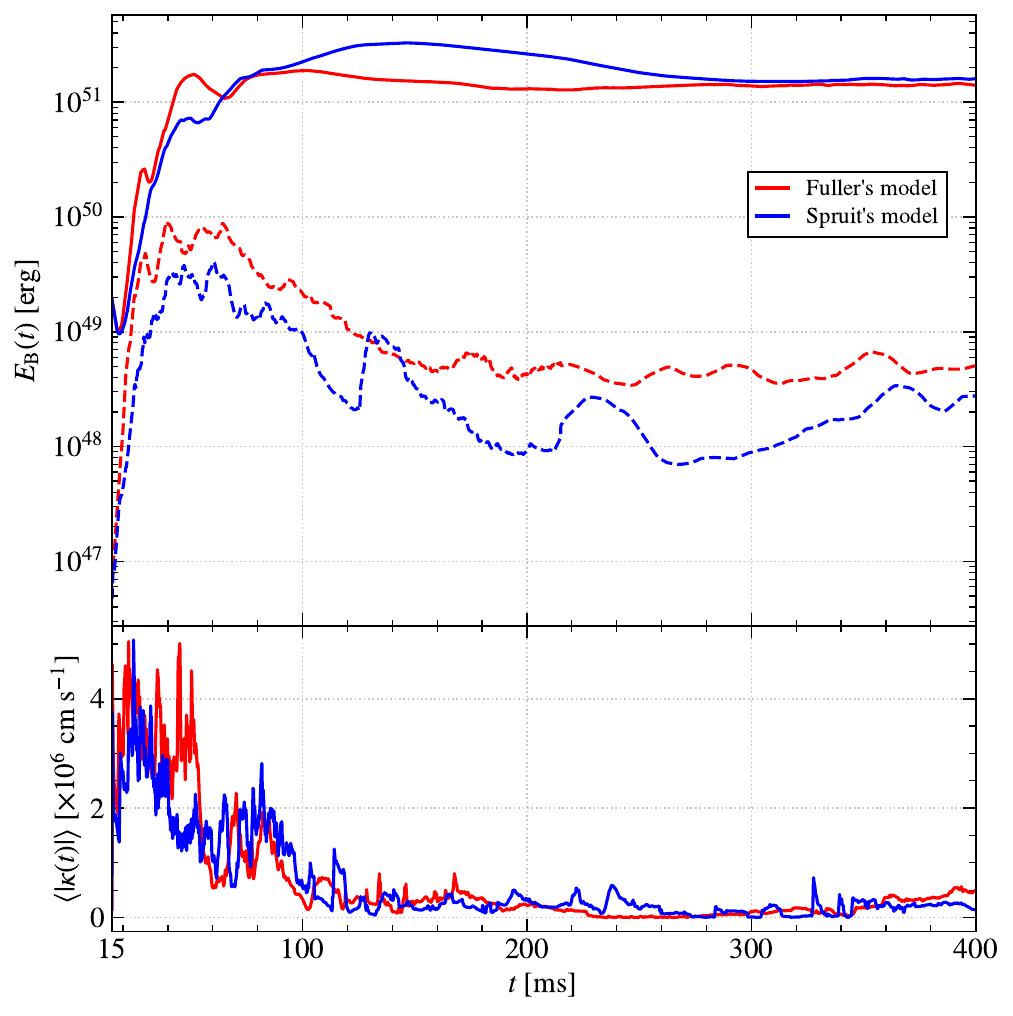}
    \caption{
    Same as Fig. \ref{fig:fig2}, but comparing Spruit model~\citep{Spruit2002} (blue lines)
    and Fuller's model (red lines) used in main text.
    We assume $\xi_{\rm TS} = 1$ and $\chi_{\rm TS} = 1$ in both cases.
    }
    \label{fig:spruit_energy}
\end{figure}

\subsection{Uncertainties in the transport model}\label{sec:Spruit_TS}
Another uncertainty lies in the prescription for the saturation of the TS dynamo, which modifies the saturation condition entering Eq.~\eqref{eq:kappa_tar} and hence feeds back onto the subgrid model.
In Sec.~\ref{sec:diff_chi}, we estimated the range of $\chi_{\rm TS}$ required to mimic variations in the saturated dynamo coefficient $\kappa_{\rm TS}$, including the uncertainty associated with the original model \citep{Spruit2002}. Here, we instead explicitly simulate this model prescription by adopting both the modified saturation condition and the corresponding $\kappa_{\rm TS}$.

More precisely, \citet{Spruit2002} proposed a different saturation prescription in which Tayler-generated turbulent motions provide an effective magnetic diffusivity that damps the large-scale toroidal field on a timescale comparable to the linear TI growth time, with saturation obtained by balancing this decay against shear amplification. This differs from the prescription of \citet{Fuller2019}, in which the TI transfers magnetic energy to the perturbed field, whose nonlinear cascade and dissipation balance the instability growth, while the equilibrium field strength is set by balancing turbulent dissipation of the perturbations against magnetic-energy injection by differential rotation.

Assuming $\xi_{\rm TS}=1$ and $\chi_{\rm TS}=1$, and the same constraints of 
$\omega_A \ll  \Omega \ll N_{\rm BV}$ and neutrino viscosity condition (Eq.~\eqref{eqn:TS_neutrino_cond}), 
we run an additional simulation by adopting  Eq.~\eqref{eq:kappa_spruit} and 
replacing the saturated condition of
Eq.~\eqref{eq:BrBphi_sat} to obtain:
\begin{equation}
\left|\frac{B^{R}}{R B^{\phi}}\right|_{\rm sat}=
\frac{\omega_{A,\mathrm{sat}}^{\mathrm{Spruit}}}{N_{\mathrm{BV}}}
=
\frac{R\,|\partial_R\Omega|\,|\Omega|}{N_{\mathrm{BV}}^2}.
\end{equation}

Figure~\ref{fig:omega_chi_comp} 
shows weaker magnetic-field growth before global saturation and a lower saturated poloidal magnetic energy than in the Fuller prescription. Nevertheless, toward the late-time evolution, at $\bar t\gtrsim300~{\rm ms}$, the two prescriptions converge to very similar total toroidal and poloidal magnetic energies.

Figure~\ref{fig:spruit_am} further shows that the overall angular-momentum gain and loss regions, as well as the flattening of the differential-rotation profile, are very similar between the two models.
The two prescriptions also produce comparable angular-momentum transport strengths at most times. The differences are more noticeable during the first $\sim100~{\rm ms}$, when the Fuller prescription produces somewhat stronger transport. After $\sim300~{\rm ms}$, the Maxwell stresses and $\Delta J_R$ in the two models become very similar.

These findings show that model uncertains primarily affect the rate of transport (fast vs. slow) but not the final state, incl. ejection of disk material.
Together with the strong background dependence (rotation and thermodynamical sate) nature of the TS dynamo discussed in the main text, this leads to only a weak dependence of the secular remnant evolution on the model choice, pointing towards the robustness of TS-dynamo physics in postmerger remnant, 
secular AM transport and magnetic field amplification.

\subsection{Uncertainty in the neutrino-viscosity constraint in merger remnants}
\label{sec:uncertain_nu_vis}

All simulations presented in this work impose the neutrino-viscosity constraint of Eq.~\eqref{eqn:TS_neutrino_cond} from \citet{Reboul-Salze2025}. Since the neutrino-opaque remnant core has a large neutrino shear viscosity, $\nu_{\nu}$, which can strongly suppress short-wavelength TI modes. For postmerger conditions, $\nu_{\nu}\sim10^{7}$--$10^{8},{\rm cm^{2},s^{-1}}$, while the magnetic diffusivity, $\eta_{\rm mag}$, is expected to be very small, giving $\mathrm{Pm}\equiv\nu_{\nu}/\eta_{\rm mag}\sim10^{12}$. Conversely, neutrino-mediated diffusion 
$D_\nu$ is much faster than momentum diffusion, with $\mathrm{Pr}\equiv\nu_{\nu}/D_{\nu}\sim10^{-3}$--$10^{-2}$ \citep{Reboul-Salze2025}. 

To approximately account for neutrino-viscous suppression, \citet{Reboul-Salze2025} assumed that viscous damping remains slower than the TI growth rate. Combined with Eq.~\eqref{eq:k_R_gt_k_N}, this gives
\begin{equation}
\nu_{\nu}k_R^2
\lesssim
\sigma_{\rm TI}
\simeq
\frac{\omega_A^2}{|\Omega|},
\label{eq:Reboul_nu_nukR}
\end{equation}
and hence the critical toroidal field in Eq.~\eqref{eqn:TS_neutrino_cond}, below which our TS subgrid model is deactivated. This provides a conservative estimate by treating viscosity as a damping process acting on the TI mode.

However, the revised linear analysis of \citet{Skoutnev2024} shows that the rapidly rotating, diffusive TI cannot generally be reduced to this single damping condition. Viscosity modifies the magnetostrophic eigenmodes and introduces a viscous branch whose growth peaks near
\begin{equation}
k_{\nu}\simeq
\left(\frac{2|\Omega|}{\nu_{\nu}}\right)^{1/2},
\qquad
\gamma_{\rm TI}(k_{\nu})
\simeq
\frac{\omega_A^2}{4|\Omega|}.
\end{equation}
In the large-$\mathrm{Pm}$ limit, when buoyancy is sufficiently weak at $k_{\nu}$, the survival of this branch is instead limited by magnetic diffusion,
\begin{equation}
\frac{\omega_A}{2|\Omega|}
\gtrsim
\mathrm{Pm}^{-1/2}
=
\left(\frac{\eta_{\rm mag}}{\nu_{\nu}}\right)^{1/2}.
\end{equation}

Moreover, neutrino-mediated diffusion can also weaken the buoyant restoring force and introduce a characteristic diffusive scale
\begin{equation}
k_{D_{\nu}}
\simeq
\left(
\frac{k_{\theta}^{2}N_{\rm BV}^{2}}
{2|\Omega|D_{\nu}}
\right)^{1/4},
\end{equation}
where 
$k_{\theta}$ is horizontal/transverse wavenumber of the TI.
The ordering between $k_{D_{\nu}}$ and $k_{\nu}$ depends on the local values of $N_{\rm BV}$, $|\Omega|$, $\nu_{\nu}$, and $D_{\nu}$, rather than on $\mathrm{Pr}$ alone. The relevant TI branch and its stability therefore depend jointly on $\eta_{\rm mag}$ and $D_{\nu}$ and $\nu_{\nu}$~\citep{Skoutnev2024}.

Therefore, the physical TI stability boundary in the expected high-$\mathrm{Pm}$, low-$\mathrm{Pr}$ regime cannot generally be reduced to a simple
$B^{\phi}_{\rm crit}(N_{\rm BV},|\Omega|,\nu_{\nu})$ used in Eq.~\eqref{eqn:TS_neutrino_cond}. 

Our simulations evolve ideal GRMHD with moment-based radiation and therefore cannot self-consistently provide the physical $\eta_{\rm mag}$, $D_\nu$ as well as $\nu_\nu$ to 
evaluate a full diffusive TI dispersion relation.
Therefore, we retain the simplified viscosity constraint as a conservative activation constraint rather than a precise linear-stability boundary. 
Given the expected hierarchy $\eta_{\rm mag}\ll\nu_{\nu}\ll D_{\nu}$, Eq.~\eqref{eqn:TS_neutrino_cond} can be substantially more restrictive than the diffusive conditions and, therefore, our simulations may underestimate the active regions of TS-dynamo. Future work will be required to provide a more self-consistent treatment.

\bibliography{aeireferences_harry,Elias}
\bibliographystyle{aasjournalv7.1}

\end{document}